\documentclass[aps,prd,twocolumn,showpacs,preprintnumbers,amsmath,amssymb]{revtex4-1}
\usepackage{graphicx}
\usepackage[usenames,dvipsnames]{xcolor}
\usepackage{braket}
\usepackage{lineno}
\usepackage{multirow}
\begin{document}
\title{Universal features of the Abelian Polyakov loop in 1+1 dimension}
\author{J. Unmuth-Yockey$^1$}
\email{jfunmuth@syr.edu}
\author{Jin Zhang$^{2}$}
\email{jzhan039@ucr.edu}
\author{A. Bazavov$^3$}
\author{Y. Meurice$^4$}
\author{S.-W. Tsai$^2$}
\affiliation{$^1$Department of Physics, Syracuse University, Syracuse, New York 13244, USA}
\affiliation{$^2$ Department of Physics and Astronomy, University of California, Riverside, CA 92521, USA}
\affiliation{$^3$Department of Computational Mathematics, Science and Engineering,
and Department of Physics and Astronomy,
Michigan State University, East Lansing, MI 48824, USA}
\affiliation{$^4$ Department of Physics and Astronomy, The University of Iowa, Iowa City, IA 52242, USA }
\definecolor{burnt}{cmyk}{0.2,0.8,1,0}
\def\lt{\lambda ^t}
\def\note{note}
\def\beq{\begin{equation}}
\def\enq{\end{equation}}
\newcommand{\Tr}{\text{Tr}}

\date{\today}
\begin{abstract}
We show that the Polyakov loop of the two-dimensional lattice Abelian Higgs model can be calculated using the tensor renormalization group approach.
We check the accuracy of the results using standard Monte Carlo simulations. 
We show that the energy gap produced by the insertion of the Polyakov loop 
obeys universal finite-size scaling which persists in the time continuum limit. We briefly discuss the relevance of these results for quantum simulations. 

\end{abstract}

\maketitle
\section{Introduction}

Two-dimensional gauge models have played an important role in our understanding of four-dimensional quantum chromodynamics (QCD). 
They appear prominently in several of the Coleman's Erice lectures \cite{coleman} and provide non-trivial model calculations for questions related to confinement, topology and symmetry breaking. For these reasons, they are often the first targets when new methods are developed. 
There has been a recent interest in using controlled quantum systems to perform calculations in lattice gauge theories. 
The methods used include cold atoms in optical lattices, trapped ions, and state of the art quantum computers.  
Recent efforts have been focused on  the Schwinger model \cite{martinez2016real,PhysRevLett.109.175302,Kasper:2017,Klco:2018kyo} and its scalar counterpart the two-dimensional Abelian Higgs model \cite{kuno2015real,kuno2017quantum,Cuadra2017}.

In recent years, the tensor renormalization group method (TRG) has been used to reformulate spin and gauge models with compact field variables into models of discrete integer (or half-integer) fields  \cite{PhysRevB.86.045139,prb87,prd88,prd89,pre89}. This reformulation uses discrete character expansions which are suitable for quantum computations and can also be used 
for sampling purposes \cite{PhysRevD.92.114508}. The computation of the tensors involves integration over the field variables and is manifestly gauge invariant. The TRG has been used to reformulate the 2D Abelian Higgs model and find approximations suitable to implement the model on optical lattices  \cite{PhysRevD.92.076003}. Recently developed experimental methods involving Rydberg atoms \cite{Zeiher2016} have been exploited to propose 
realistic implementations of the model on a physical ladder \cite{Zhang18}.

In order to test the ladder implementation, we proposed to measure the Polyakov loop \cite{Zhang18} for the 2D Abelian Higgs model. In this process we found remarkable finite-size scaling (FSS) properties that to the best of our knowledge have never been reported. In this article, we 
describe these calculations and the interpretation of the results. 

The paper is structured as follows: In Sec. \ref{sec:abh_pl} the reformulation of the model is briefly reviewed, and the Polyakov loop is introduced in terms of the reformulated variables. We emphasize that the Brout-Englert-Higgs mode is taken to be infinitely massive and that we are left with the compact Nambu-Goldstone modes and compact gauge fields. The model that we are considering could be called 
``compact scalar electrodynamics''.

In Sec. \ref{sec:num-calc-pl} numerical calculations in the relativistic Lagrangian formulation where space and Euclidean time are on the same footing are 
reported. We first show that the TRG method based on coarse graining and the standard Monte Carlo (MC) sampling on the original fields are in good numerical agreement. We then show that the Polyakov loop defines an energy gap $\Delta E$ that can be extracted from lattice configurations 
with different temporal lengths.  We report on the FSS of this energy gap and present results across a range of spatial sizes and gauge couplings.

In Sec. \ref{sec:pl-time-con} we review the continuous-time limit for this model.  We then relate this continuous-time limit in the field-variables representation to a Hamiltonian in the charge-variables representation.  Next we derive the continuous-time limit of the Polaykov loop and show that its insertion can be realized by a local modification of the original Hamiltonian.  It is here that we show that the FSS observed in the isotropic coupling case survives the continuous-time limit of this model and exhibits similar data-collapse.  Finally, we consider the limit of zero gauge coupling where the model reduces to that of the O(2) spin model, and we give justification for the accuracy of our results.

In Sec. \ref{sec:01bc} we relate a special choice of boundary conditions in this model to the inclusion of the Polyakov loop into the system.  We find this special boundary condition allows us to probe the non-zero charge sectors of the theory.  It is discovered that the energy gap extracted from this special choice of boundary conditions exhibits similar FSS compared to the energy gap extracted from the Polyakov loop.  Again, this FSS is found to persist into the continuous-time limit.
Finally in Sec. \ref{sec:conclusion} we give a summary and concluding remarks about the pertinence of this study to the possibility of quantum simulating the 2D Abelian Higgs model using a ladder set-up with cold atoms in an optical lattice.

\section{The Abelian Higgs model and the Polyakov loop}
\label{sec:abh_pl}

\subsection{The model and its gauge-invariant reformulation}
\label{subsec:reform}

In this paper we consider the compact Abelian Higgs model (scalar electrodynamics) with the Higgs mode frozen to unity in 1+1 Euclidean spacetime dimensions using a lattice discretization.  The lattice has spatial and temporal extents $N_s$, and $N_\tau$, respectively.  We used a variety of boundary conditions including periodic (PBC), open boundary conditions (OBC) in space, and more exotic boundary conditions.  We will mention which type was used when necessary; however, typically calculations done in the discrete Lagrangian set-up were done with PBC, while calculations done in the continuous-time limit were done with OBC.  This model has been introduced before in Ref. \cite{PhysRevD.92.076003}.
The action for this model is
\begin{align}
	S = &-\beta_{pl.}\sum_x\sum_{\nu<\mu}{\rm Re}{\rm Tr}\left[U_{x,\mu\nu}\right] \nonumber
\\
&-\kappa\sum_x\sum_{\nu=1}^{2}
\left[\phi_x^\dagger U_{x,\nu}\phi_{x+\hat\nu}+
\phi_{x+\hat\nu}^\dagger U^\dagger_{x,\nu}\phi_x
\right].
\end{align}
with $U_{x,\mu\nu} = e^{i (A_{x,\mu} + A_{x+\mu,\nu} - A_{x+\nu, \mu} - A_{x, \nu})}$, $U_{x, \mu} = e^{i A_{x,\mu}}$, and $\phi_{x} = e^{i \theta_{x}}$. The gauge coupling enters into $\beta_{pl} = 1/g^{2}$.  The coupling $\kappa$ controls the scalar-field hopping between nearest-neighbor sites.
The partition function for this model is
\begin{equation}
	Z = \int \mathcal{D}[\phi^{\dagger}] \mathcal{D}[\phi] \mathcal{D}[U]
    e^{-S}
\end{equation}
with $\mathcal{D}[\phi] = \prod_{x} d\phi_{x}$, and similarly for the gauge field integration.
 Because of the compact variables of integration in the original formulation ($\theta_x$ and $A_{x,\hat\nu}$) the Boltzmann weights can be expanded using Fourier analysis \cite{RevModPhys.52.453}.
From these expansions the partition function can be rewritten exactly in a gauge-invariant way by integrating out the $\theta$ and $A$ fields.  One is left with only integer fields on the links and plaquettes which can be further simplified to integer fields only living on the plaquettes.  The partition function can then be written as
\begin{equation}
	Z = \sum_{\{ m \}} \left( \prod_{x, \nu < \mu} I_{m}(\beta_{pl}) \right)
    \left( \prod_{x, \nu} I_{m-m'}(2\kappa) \right).
\end{equation}

In what follows we always normalize the Bessel functions by their smallest order, i.e. we use the following definitions:
$t_{n}(z)\equiv I_{n}(z)/{I_{0}(z)}$,
$t_n(0)=\delta_{n,0}$.
For $0 < z < \infty$ we have
$1>t_0(z)>t_1(z)>t_2(z)>\dots>0$.  In addition for large $z$,
$t_n(z)\simeq 1-n^2/(2z)$ and for small $z$, $t_{n}(z) \simeq z^{n}/(2^{n} n!)$.

\subsection{The Polyakov loop}
\label{subsec:ploop}

The Polyakov loop, $P$, is a specific instance of the Wilson loop.  The later is defined by closed loops built out of gauge fields, and is gauge invariant.  The Polyakov loop is a Wilson loop which wraps around the (closed) temporal direction (with PBC) making it non-contractible.  The Polyakov loop is an order parameter for confinement/deconfinement transitions in gauge theories.  In particular it monitors the center symmetry of the gauge group, and the screening of a static test charge by the gauge field.

The Polyakov loop is related to the free energy induced by the inclusion of the static charge by \cite{Montvay:1994cy}
\begin{equation}
	\exp[-F/kT] \propto \langle P \rangle .
\end{equation}
As defined above, the Polyakov loop has the form
\begin{equation}
	P = \prod_{n=0}^{N_{\tau}-1} U_{x^{*}+n\hat{\tau}, \hat{\tau}}
\end{equation}
in the Abelian Higgs model considered here, which is a loop along a single space-slice.
With PBC, the insertion of the Polyakov  loop into the system forces a scalar current in the opposite direction in order to lower the system energy;
however, the cost for the current to run the length of $N_{\tau}$ is controlled by the hopping parameter coupling and the length of $N_{\tau}$, and this cost must be overcome for the Polaykov loop expectation value to be nonzero.

 We can re-write the Polyakov loop in terms of the gauge-invariant variables of Sec. \ref{subsec:reform}.  Consider the expectation value of the Polyakov loop,
\begin{equation}
	\label{eq:ploop}
	\langle P \rangle = \frac{1}{Z} \int \mathcal{D}[\phi^{\dagger}] 
    \mathcal{D}[\phi] \mathcal{D}[U]
    \left( \prod_{n=0}^{N_{\tau}-1} U_{x^{*}+n\hat{\tau},\hat{\tau}} \right)
    e^{-S}
\end{equation}
with $x^{*}$ a single specific spatial site.
Using the expansions from the gauge-invariant reformulation from before we pick up new link integrals on the links which contain the additional $U$ variables from the Polyakov loop,
\begin{equation}
	\int \frac{\theta_{x}}{2 \pi} e^{i(n - m_{r} + m_{l} + 1)\theta_{x}} = \delta_{n, m_{r}-m_{l}-1}.
\end{equation}
Here the subscripts $l$ and $r$ denote the ``left'' and ``right'' plaquette quantum numbers, respectively, to the vertical (temporal) link in question.  This shifts the difference in $m$s by one at the links which contain the Polyakov loop, but all the other links remain the same.  Now we can write the expectation value as,
\begin{align}
	\langle P \rangle = \frac{1}{Z} \sum_{\{ m \}} &\left[ \prod_{x, \nu < \mu}
    t_{m}(\beta_{pl}) \right] \left[ \prod_{x, \nu}
    t_{m-m'}(2 \kappa) \right] \times \nonumber \\
    & \left[ \prod_{n=0}^{N_{\tau}-1}
    \frac{t_{m-m'-1}(2 \kappa)}{t_{m-m'}(2 \kappa)} \right].
\end{align}
Here the last product is over those links from Eq. \eqref{eq:ploop} that are included in the Polyakov loop.  This allows us to identify the Polyakov loop in terms of the new variables as,
\begin{equation}
	P = \prod_{n=0}^{N_{\tau}-1}
    \frac{t_{m-m'-1}(2 \kappa)}{t_{m-m'}(2 \kappa)}.
\end{equation}

\section{Isotropic calculations of the Polyakov loop}
\label{sec:num-calc-pl}

\subsection{MC and TRG}
In this section we explore
the construction and implementation of the Polyakov loop in terms of the reformulated $m$ variables using the TRG.  In order to check our work and results we compared with traditional MC methods.  The MC algorithms used were the same as those implemented in Ref. \cite{PhysRevD.92.076003}.  For the TRG calculations we used the higher-order tensor renormalization group (HOTRG) method.  We typically used a bond dimension, $D_{\text{bond}}$, of 41 states, and in some cases, 51 states were used to assess the fluctuations.  The tensor used in the calculation was constructed in a similar way to that described in Ref. \cite{PhysRevD.92.076003}.  The main tensor was constructed from a $B$ tensor, and the four legs of that tensor were contracted with the square-root, or Cholesky decomposition, of the $A$ tensor,
\begin{equation}
	t_{m-m'}(2 \kappa) \equiv A_{m m'}(2 \kappa) = L_{m \alpha} L^{\dagger}_{\alpha m'}(2 \kappa).
\end{equation}
This is possible when the eigenvalues of this matrix are positive which, in all cases considered here, they are.  One can then combine the $L$ matrix here with the $B$ tensor from Ref. \cite{PhysRevD.92.076003}.  The transfer matrix can be constructed by blocking along a time-slice with the tensor described above.

To find the expectation value of the Polaykov loop, we considered another transfer matrix, which possesses an impure temporal link: that link which is shifted by one due to the addition of the line of $U$s from the Polyakov loop.  This can be accomplished by inserting a matrix into the end of the blocked time-slice such that when the ends are closed (due to PBC) the resulting $A$ matrix which would be completed by this contraction is instead a matrix of the form,
\begin{equation}
	t_{m - m' - 1}(2 \kappa) \equiv \tilde{A}_{m m'}(2 \kappa).
\end{equation}
Since this is a transfer matrix, this impure matrix will appear $N_{\tau}$ times in the product thus defining the Polyakov loop of length $N_{\tau}$.

\begin{figure}[t]
  \includegraphics[width=0.49\textwidth]{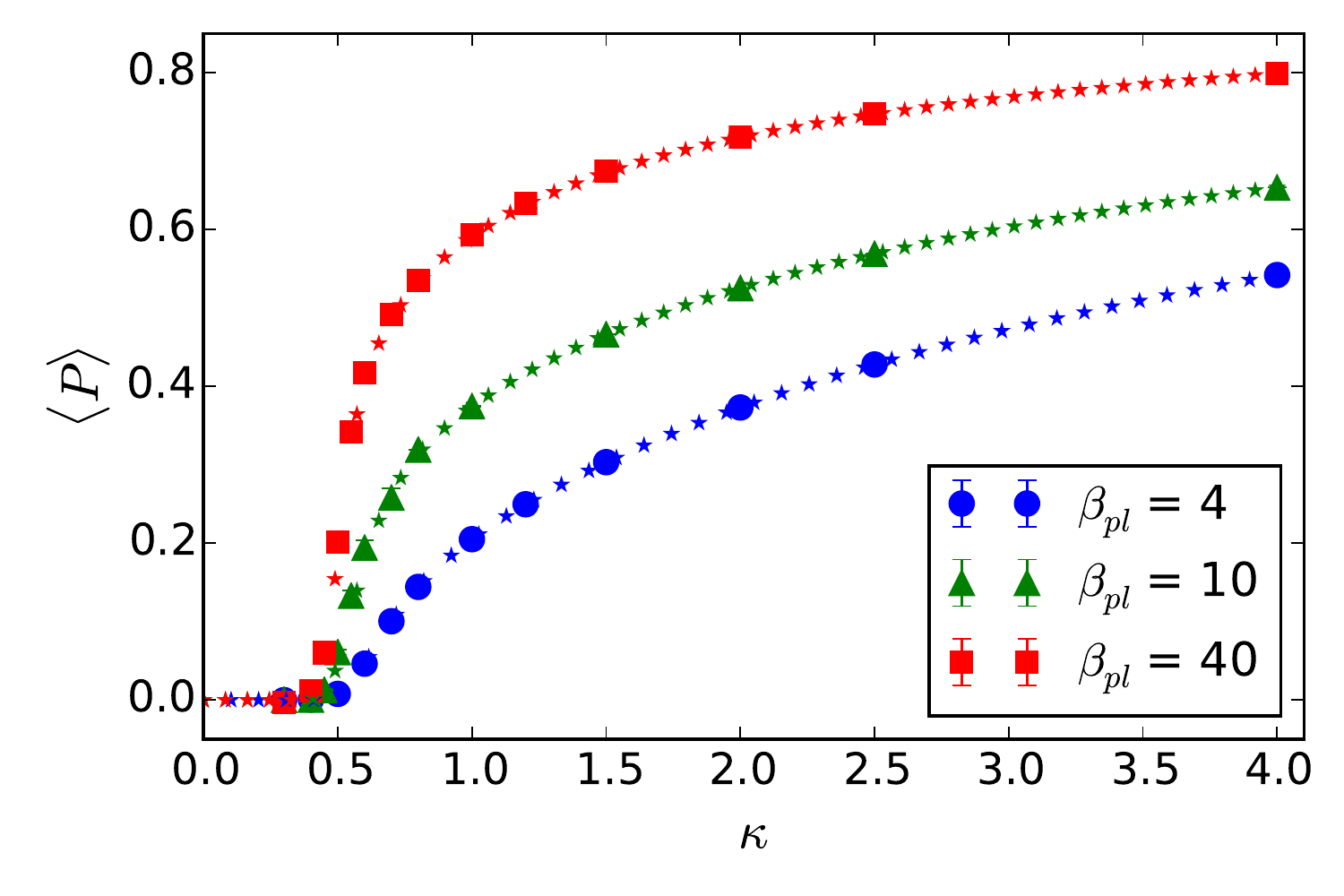}
  \caption{Comparison between HOTRG and MC for a range of $\kappa$ and $\beta_{pl}$ values.  Here the HOTRG data appears as the star markers, while the MC appears as the square, triangle, and circle markers with error-bars.  This was done on a lattice of size $N_{s} = N_{\tau} = 16$.}
  \label{fig:ploop_mc_trg_comp}
\end{figure}

\begin{figure}[t]
	\includegraphics[width=0.49\textwidth]{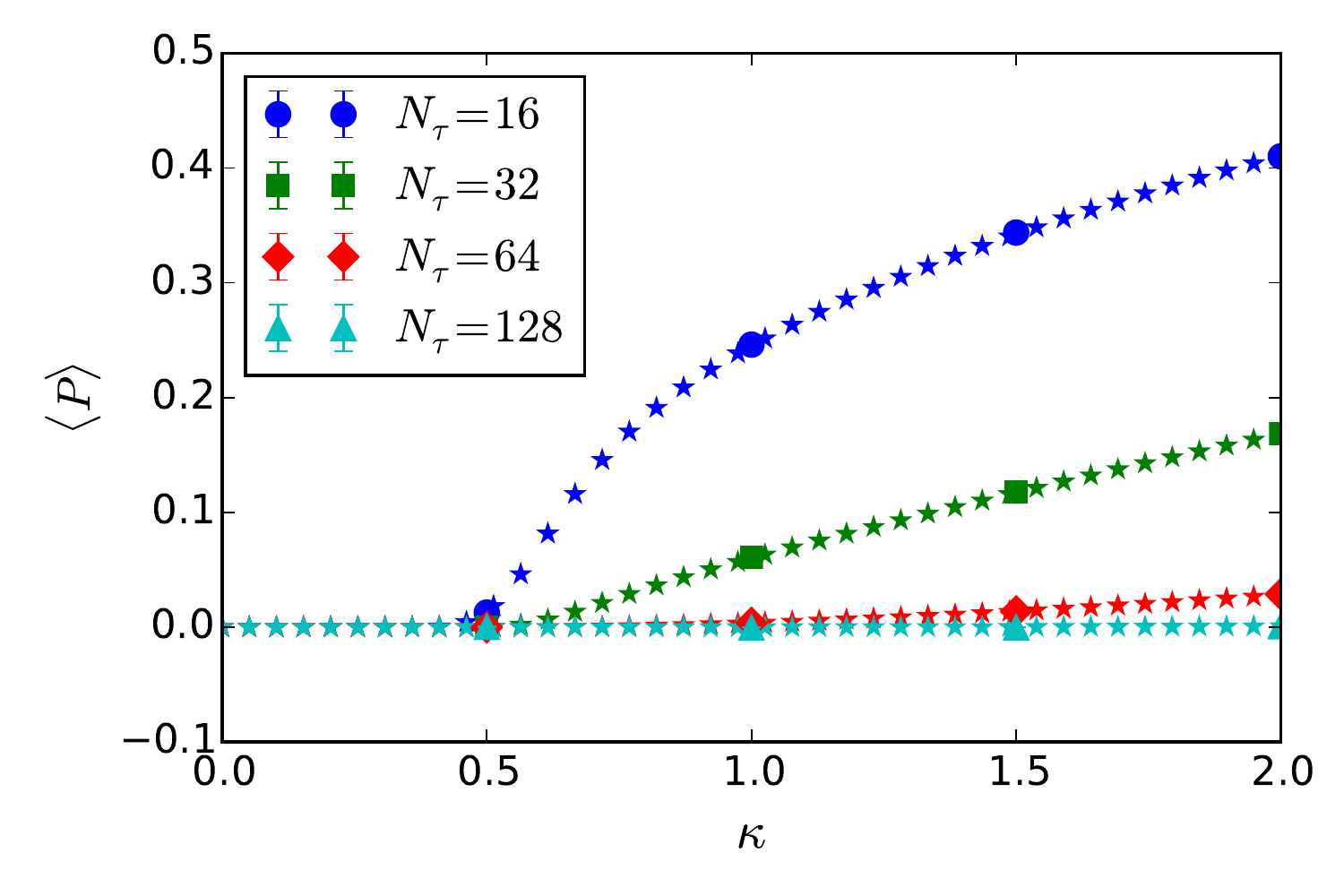}
	\caption{HOTRG and MC data with fixed spatial length and various temporal lengths.  Data like this was used to find the decay in the temporal direction of the lattice.  It was found to decay exponentially for large enough temporal lengths.  Here the stars are the HOTRG data while the squares, circles, diamonds, and triangles are the MC data with error bars.  This was done on a $N_{s} = 16$ lattice with $\beta_{pl} = 5$, and $D_{\text{bond}} = 41$.}
    \label{fig:ntau_scaling}
\end{figure}

A comparison between the HOTRG method and MC for computing the Polyakov loop can be seen in Fig. \ref{fig:ploop_mc_trg_comp}.  In this figure, we keep a fixed spatial and temporal extent with PBC, and varied the gauge coupling along a fixed range of $\kappa$ values.  This can be compared with calculations done in Ref. \cite{doi:10.1142/S0217751X08041281}.  Another example of the comparison between the two can be seen in Table \ref{tab:mc_trg},
where we varied the temporal extent of the lattice at fixed spatial size for a variety of $\kappa$ values.  This data was a catalyst for the study of the scaling of the free energy, or energy gap, of the static charge inserted into the system, via, the Polyakov loop.  Fig. \ref{fig:ntau_scaling} shows the data from Table \ref{tab:mc_trg} of the comparison between MC and HOTRG when varying $N_{\tau}$.  Again, this was done using PBC.  Overall we find good agreement between the two methods.

\begin{center}
	\begin{table}
	\begin{tabular}{|c | c | c | c | c|}
    \hline
        & $\kappa = 0.5$ & $\kappa = 1$ & $\kappa = 1.5$ & $\kappa = 2$ \\
        \hline
        \multirow{2}{*}{$N_{\tau} = 16$} & 0.0136(1) & 0.2451(1) & 0.3424(1) & 0.4102(1) \\
        & 0.013(1) & 0.2461(6) & 0.3438(5) & 0.4102(7) \\
        \hline
        \multirow{2}{*}{$N_{\tau} = 32$} & 1.84(2)e-4 & 0.0601(1) &  0.11725(4) & 0.16828(3) \\
        & 2(4)e-4 & 0.0606(4) & 0.1175(4) & 0.1687(4) \\
        \hline
        \multirow{2}{*}{$N_{\tau} = 64$} & 3.4(1)e-8 & 0.00361(1) & 0.01374(1) & 0.02832(2) \\
        & -1(3)e-4 & 0.0037(4) & 0.0137(3) & 0.0288(4) \\
        \hline
        \multirow{2}{*}{$N_{\tau} = 128$} & 1.15(4)e-15 & 1.3(1)e-5 & 1.89(1)e-4 & 8.02(1)e-4 \\
        & 3(3)e-4 & 4(2)e-4 & 3(2)e-4 & 4(2)e-4 \\
        \hline
	\end{tabular}
    \caption{A table comparing MC and HOTRG values for the Polyakov loop.  The upper values in each cell are the values gotten from HOTRG blocking, while the lower values are those from MC calculations and their jack-knife errors.  These were generated at $\beta_{pl} = 5$ with $N_{s} = 16$.  Here the number of states kept by the tensor truncation was $D_{\text{bond}} = 41$, and $D_{\text{bond}} = 33$ was used to estimate the errors.}
    \label{tab:mc_trg}
    \end{table}
\end{center}

\subsection{The energy gap}
\label{subsec:deltae}

\begin{figure}
  \includegraphics[width=0.45\textwidth]{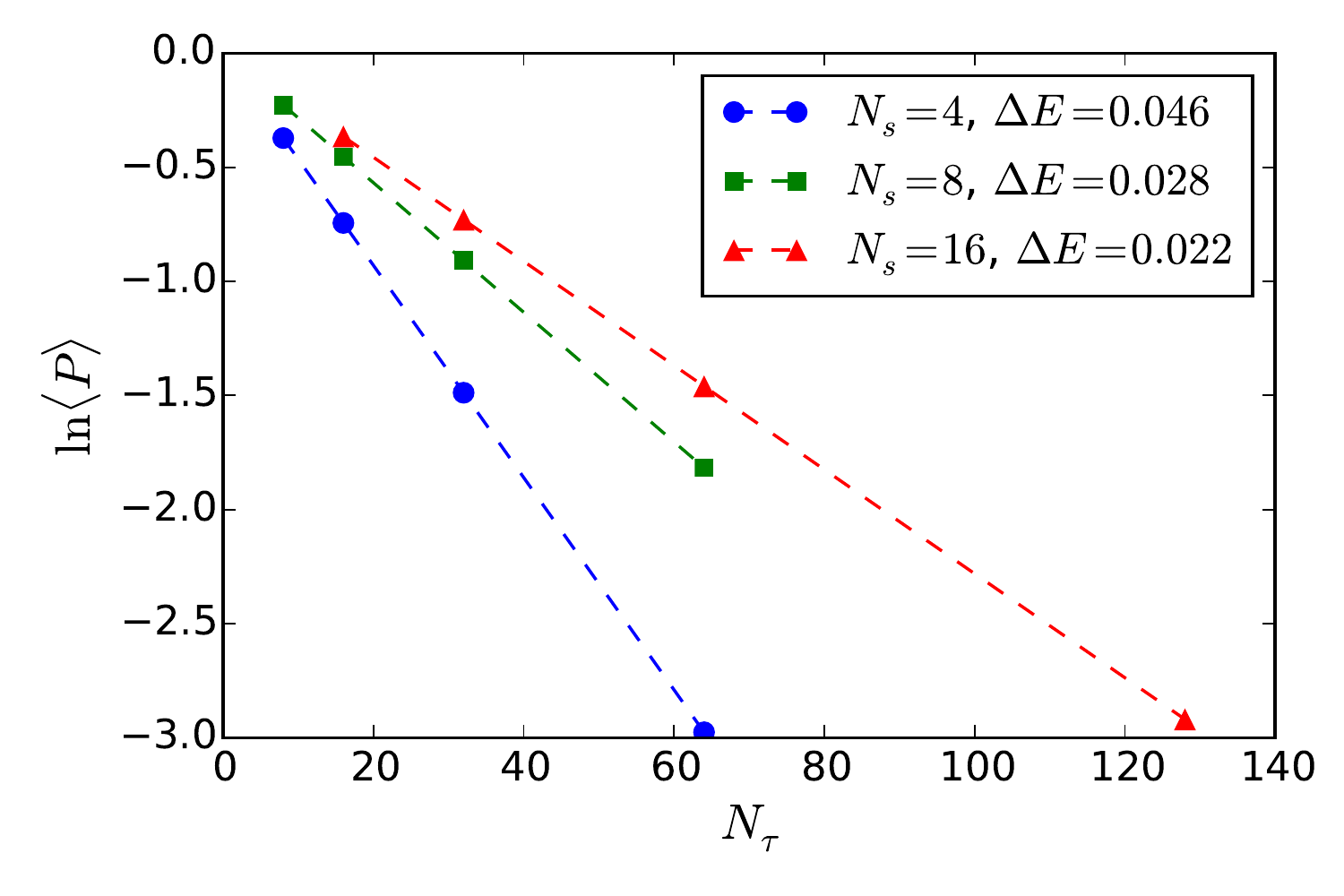}
  \caption{The energy gap for different spatial sizes with $\kappa = 1.6$ and $\beta_{pl} = 44$.  In general the slope depends on $N_s$, $\beta_{pl}$, and $\kappa$.  
The dashed lines are \emph{not} fits, merely lines connected between dots to guide the eye.  $D_{\text{bond}} = 41$ was used for these HOTRG calculations.}
  \label{fig:slopes}
\end{figure}

\begin{figure}[t]
	\includegraphics[width=0.49\textwidth]{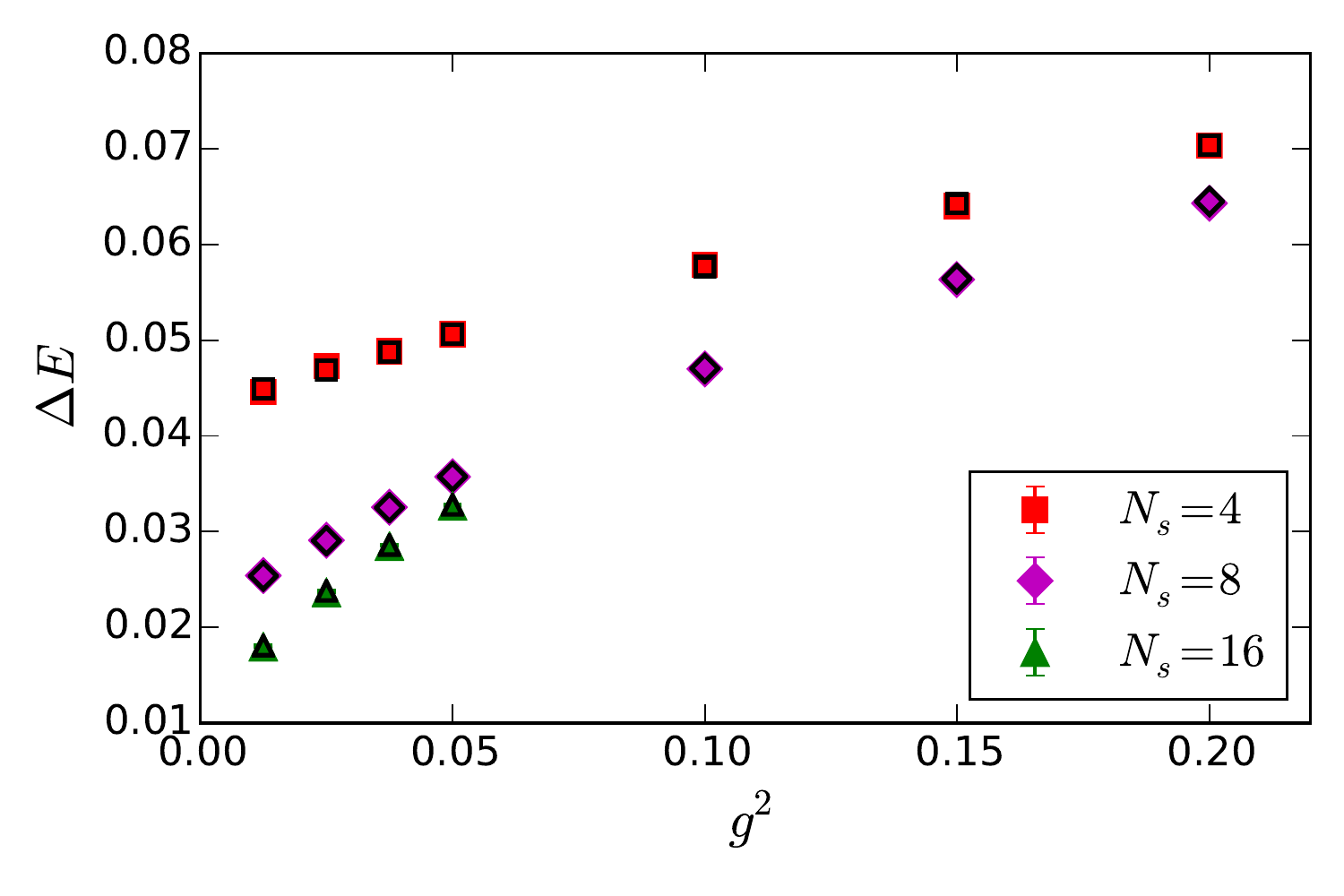}
    \caption{This figure shows a comparison between MC and HOTRG for values of $\Delta E$ calculated at a fixed value of $\kappa = 1.6$.  Here $D_{\text{bond}} = 41$ was used for the tensor truncation.  The solid markers are the MC data, while the black empty markers are the HOTRG data.}
    \label{fig:deltae-mc-trg}
\end{figure}

Consider the Polyakov loop in terms of the ratio of two partition functions: one with the inclusion of the static charge, and the other without,
\begin{equation}
	\langle P \rangle = \frac{\tilde{Z}}{Z}.
\end{equation}
In practice, say for MC calculations, one can sum over the possible locations of $x^{*}$ in Eq. \eqref{eq:ploop} and divide by $N_{s}$.  Due to translation invariance one just recovers $N_{s}$ copies of the same number.
By re-writing the partition function as a trace over products of transfer matrices, we can expose the dependence on $N_{\tau}$,
\begin{equation}
	\label{eq:ploop-zratio}
	\langle P \rangle = \frac{\tilde{Z}}{Z} =
    \frac{\Tr[\tilde{\mathbb{T}}^{N_{\tau}}]}
    		{\Tr[\mathbb{T}^{N_{\tau}}]} =
            \frac{\sum_{i=0}^{N} \tilde{\lambda}^{N_{\tau}}_{i}}
            {\sum_{i=0}^{N}\lambda^{N_{\tau}}_{i}}
\end{equation}
where in the last step we have diagonalized the transfer matrices.  This makes it clear that in the large $N_{\tau}$ limit the Polyakov loop expectation value is dominated by the largest eigenvalues, $\tilde{\lambda}_{0}$, $\lambda_{0}$.  Thus, we find,
\begin{align}
	\log\langle P \rangle &\simeq N_{\tau}
    \log(\tilde{\lambda}_{0} / \lambda_{0}) \\
    &= - N_{\tau} \Delta E
\end{align}
with $\Delta E$ the energy gap between the ground state of the system with the static charge, and that without, for sufficiently large $N_{\tau}$.  From the previous steps we see that at sufficiently large $N_{\tau}$ (or low temperatures),
\begin{equation}
	P \simeq e^{-N_{\tau} \Delta E}.
\end{equation}
This relationship is clear in Fig. \ref{fig:slopes}, where the linear behavior is seen on a log-plot.  In addition the $y$-intercept is approximately zero indicating $N_{\tau}$ has been taken large enough.
To further test the agreement between MC and HOTRG we compared the energy gap values for a few spatial lattice sizes at fixed $\kappa$ using PBC.  Fig. \ref{fig:deltae-mc-trg} demonstrates the agreement for $\Delta E$ values.

\subsection{Universality: expectations and conjectures}
\begin{figure}
        \includegraphics[width=0.49\textwidth]{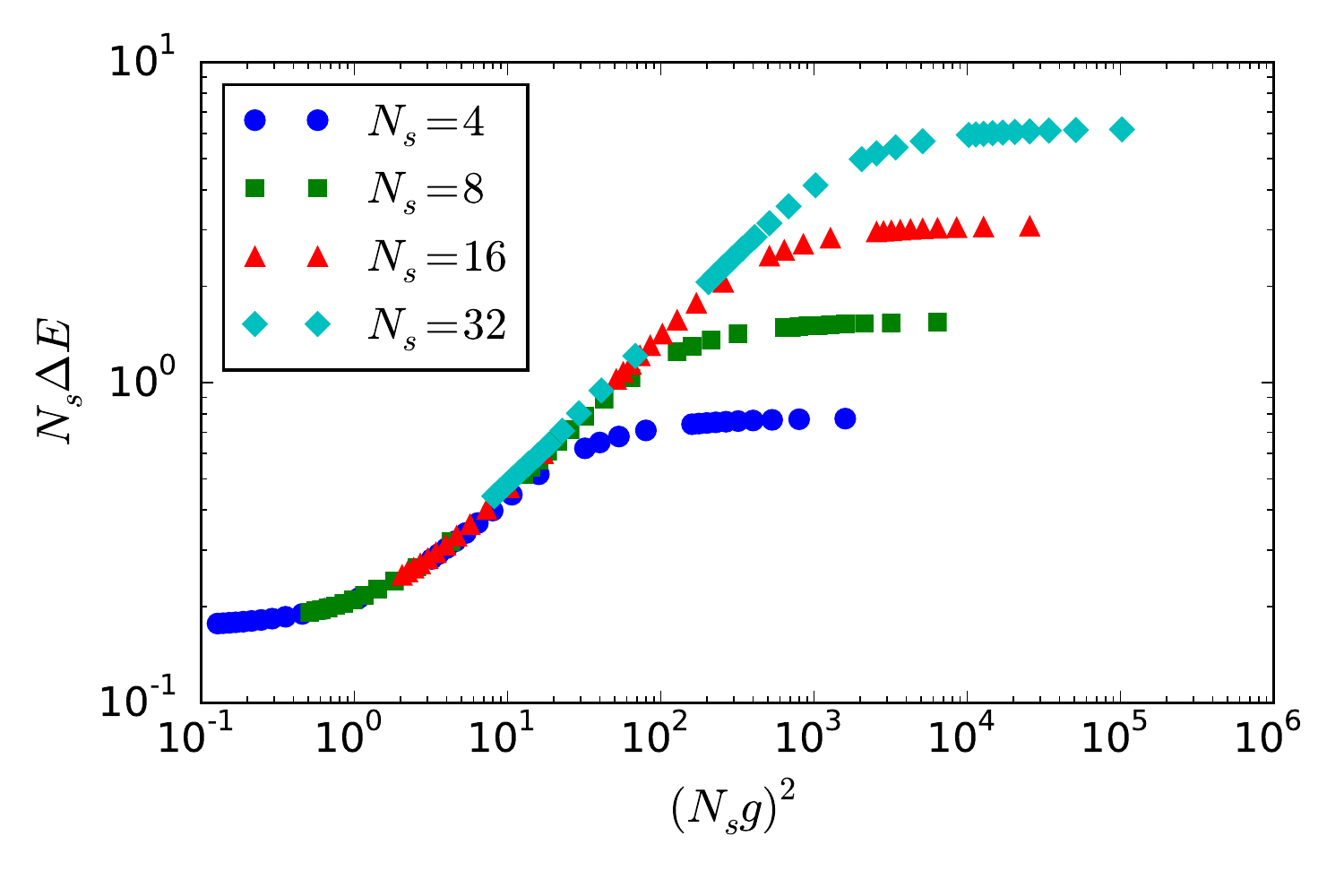}
         \caption{Data collapse across different $N_{s}$ for sufficiently small $g$, and collapse breaking across different $N_s$ at large $g$ in the case of isotropic coupling.  Here $\kappa = 1.6$, and $D_{\text{bond}} = 41$ was used in the HOTRG calculations.}
         \label{fig:iso-gap-collapse}
\end{figure}

\begin{figure}[h]
        \includegraphics[width=0.47\textwidth]{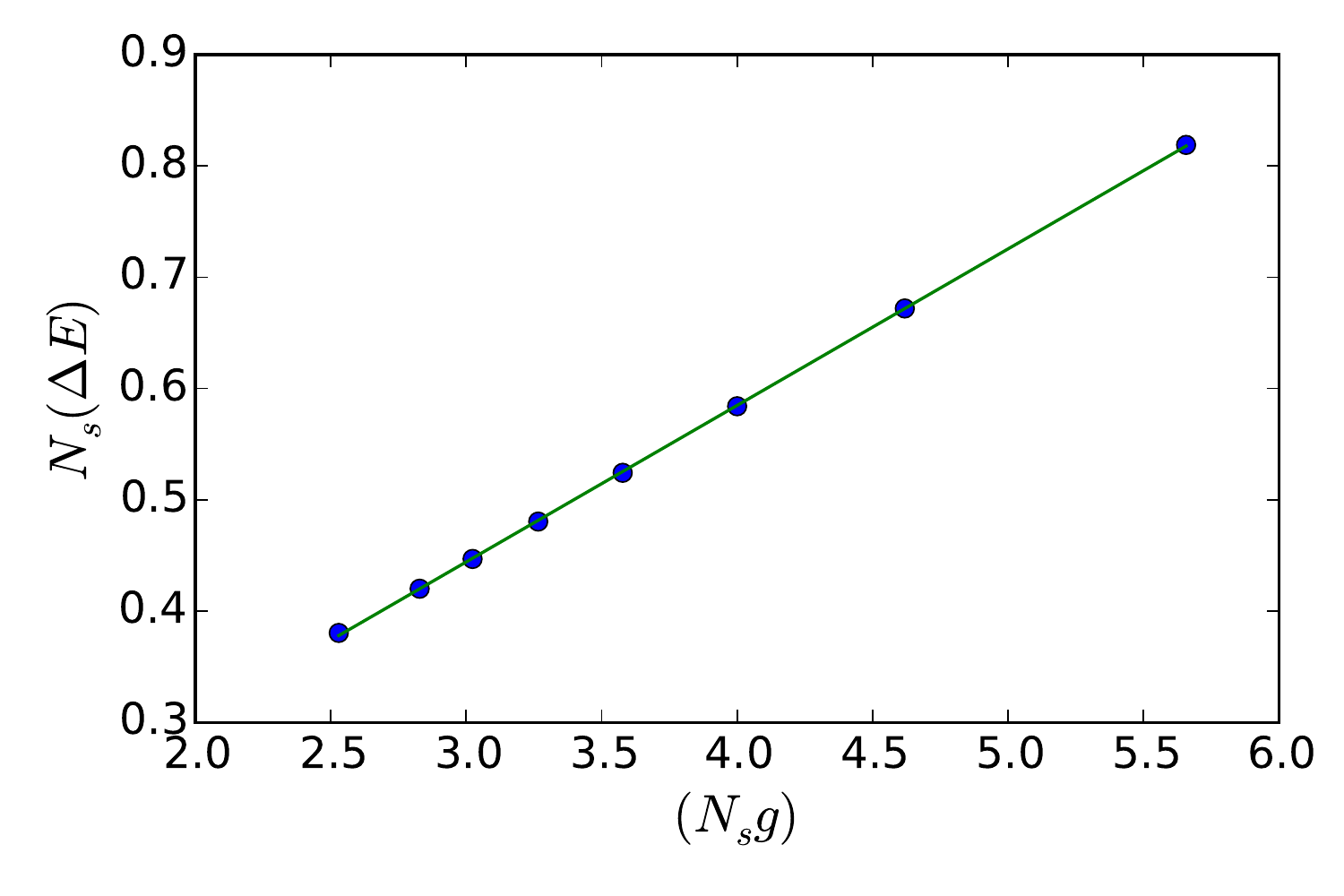}
	\caption{Data from the collapse data-set.  Here the intermediate region is plotted which exhibits a $N_{s} \Delta E \propto N_{s} g$ behavior meaning the energy gap is proportional to the gauge coupling.  The solid line is a linear fit through the data (dots).}
	\label{fig:pcol-sqrt}
\end{figure}

 In the following, we use $g^2=1/\beta_{pl}$.
 For $\kappa$ large enough, \emph{i.e.} greater than the Kosterlitz-Thouless (KT) transition value, and $g^2 N_s$ small enough, we expect that  
\begin{equation}
  \Delta E\simeq \frac{a}{N_s} + b \, g^2 N_s,
  \label{eq:gap}
\end{equation}
where $a$ and $b$ are still functions of $\kappa$.  In the limit where $g^2$ becomes zero, this is just the statement that $\Delta E$ goes to zero in the limit of large $N_s$, a consequence of the gapless KT phase at infinite volume. The guessed correction corresponds to a linear potential. If we multiply Eq. (\ref{eq:gap}) by $N_s$, then the
  right hand side depends only on $g^2N^2$. We conjecture that this feature persists beyond the lowest order approximation, namely: 
\beq
\Delta E N_s=f(g^2N^2_s). 
\enq 
Fig. \ref{fig:iso-gap-collapse}  supports this idea and shows a reasonably good data collapse across a wide range. In addition, we can observe that 
for larger $g^2 N^2_s$, $f(g^2N^2_s) \sim g N_s$, which means that in this intermediate regime, 
$\Delta E$ become approximately independent of $N_s$ and is proportional to $g$.  This intermediate region is shown in Fig. \ref{fig:pcol-sqrt} where the proportionality between $\Delta E$ and $g$ is clear, and the data is plotted against a linear fit.

However, the data collapse breaks down if we increase $g$ to large values  while keeping $N_s$ constant. 
For very large $g$ (small $\beta_{pl}$), the lowest energy state corresponds to having all plaquette quantum numbers set to zero. This is accomplished when the matter loop follows exactly the Polyakov loop in the opposite direction.  This state contributes $(t_{1}(2\kappa))^{N_{\tau}}$ to the partition function with a corresponding energy difference, $\Delta E$, of $-\ln(t_{1}(2\kappa))$ in the large $N_{\tau}$ limit.  Thus, for large values of $g$, we expect
\begin{equation}
	\Delta E \rightarrow -\ln(t_{1}(2\kappa)),
\end{equation}
independent of $N_{s}$.

\section{The Polyakov loop in the time continuum limit}
\label{sec:pl-time-con}

\subsection{The spin-1 Hamiltonian}
\label{subsec:hamil}

In Ref. \cite{PhysRevD.92.076003} the continuous-time limit for the Abelian Higgs model was taken in the field quantum number representation in the limit the Higgs quartic self-coupling goes to infinity.  We summarize the important points here.
To take the continuous time limit, one takes $\kappa_{\tau}, \beta_{pl} \rightarrow \infty$ while simultaneously taking $\kappa_{s}$, and the temporal lattice spacing, $a$, to zero such that the combinations
\begin{equation}
\label{eq:abhamil}
	U \equiv \frac{1}{\beta_{pl} a} = \frac{g^{2}}{a}, \quad
    Y \equiv \frac{1}{2 \kappa_{\tau} a}, \quad
    X \equiv \frac{2 \kappa_{s}}{a}
\end{equation}
are finite. Note that $X$ here is related to $\tilde{X}$ in Ref. \cite{PhysRevD.92.076003} by $X = \sqrt{2} \tilde{X}$.  In this limit, and using the properties of the $t_{n}(z)$ functions mentioned in Sec. \ref{subsec:reform}, the transfer matrix is close to the identity. Keeping to first-order in each coupling constant, a Hamiltonian with a three-state, or spin-1, approximation can be identified as
\begin{align}
	\label{eq:ham}
	H &= \frac{U}{2}\sum_{i=1}^{N_{s}} \left(L^z_{i}\right)^2 \nonumber \\
	&+ \frac{Y}{2} {\sum_i}  ' (L^z_{i+1} - L^z_{i})^2-
	\frac{X}{\sqrt{2}} \sum_{i=1}^{N_{s}} L^x_{i} \ ,
\end{align}
where the sum, $\sum_i '$, takes the OBC into account and includes $(L^z_1)^2 + (L^z_{N_s})^2$. 
This Hamiltonian describes a three-state system with a local Hilbert state-space of $m = \pm 1, 0$.

The first term represents the plaquette interactions.
The second term is associated with the integration of the time links. They can be interpreted as charges determined by Gauss's 
law, in other words the difference between the two plaquettes (electric field) on each side of the link.
Finally the third term is a spin-flip term, or spatial hopping term.

\subsection{The spin-$n$ Hamiltonian}
\label{subsec:gen_ham}
In order to extend and improve the study of this model in the continuous-time limit, it is advantageous to consider a spin representation of the $L^{z}$ operator greater than one.  The $X$ term in Eq. \eqref{eq:abhamil} must be modified then, since, higher spin representations of the $L^{x}$ operator do not accurately represent the genuine time continuum limit of the Abelian Higgs model.

The operator $U^x = \frac{1}{2} (U^{+} + U^-)$, with $U^{+}$ and $U^{-}$ as special types of raising and lowering operators for field quantum numbers, is the appropriate replacement for $L^{x}$ (the notation $U^{x}$ is firstly not to be confused with the parameter $U$ which parameterizes the strength of the gauge field interaction, and secondly is used to indicate that the action $U^{x}$ is similar in spirit to $L^{x}$; however, with different matrix elements). Note that $U^{\pm}$ are different from the ladder operators in the angular momentum algebra. In the basis of eigenvectors of $L^z$, applying $U^{+}$ ($U^{-}$) to them also raises (lowers) the electric field quantum number by $1$ but with all coefficients $1$,
\begin{equation}
\label{eq:ladderOp}
	U^{\pm} \ket{m} = \ket{m \pm 1}.
\end{equation}
The action of $L^{z}$ is the same, $L^z \ket{m} = m \ket{m}$.  Now the quantum Hamiltonian for the Abelian Higgs model for arbitrary spin is written
\begin{align}
	\label{eq:any-spin-ham}
	H &= \frac{U}{2}\sum_{i=1}^{N_{s}} \left(L^z_{i}\right)^2 \nonumber \\
	&+ \frac{Y}{2} {\sum_i}  ' (L^z_{i+1} - L^z_{i})^2-
	X \sum_{i=1}^{N_{s}} U^x_{i} \ ,
\end{align}
Where $X$ is the same as in Eq. \eqref{eq:ham} because of the definition of $U^{x}$.  We will still call the $2n+1$ state truncation a ``spin-$n$'' truncation for convenience, even though we no longer are using the angular momentum algebra.

\subsection{The charge representation}
By Gauss's Law, the charge (link) quantum numbers are defined as $\bar{L}_{i+\frac{1}{2}}^z = L_{i+1}^z - L_{i}^z$. This allows one to use the charge representation. The corresponding $\bar{U}^{\pm}$ have the same form. If we increase the field quantum number at site $i$ by $1$, the charge quantum number $\bar{L}_{i+\frac{1}{2}}^z$ will be decreased by $1$, but $\bar{L}_{i - \frac{1}{2}}^z$ will be increased by $1$. So the ladder operators in field representation are related with the ones in charge representation by
\begin{equation}
\label{eq:laddersrelation}
	U_i^{+} = \bar{U}_{i-\frac{1}{2}}^+ \bar{U}_{i+\frac{1}{2}}^-.
\end{equation}

With OBC, $\bar{L}^z_{\frac{1}{2}} = L^z_1$, $\bar{L}^z_{N_s + \frac{1}{2}} = -L^z_{N_s}$, we can solve for field quantum numbers in terms of charge quantum numbers

\begin{equation}
\label{eq:fieldcharge}
	L^z_i = \sum_{j = 0}^{i-1} \bar{L}^z_{j+\frac{1}{2}}.
\end{equation}
Then Eq. \eqref{eq:any-spin-ham} reads

\begin{align}
	\label{eq:any-spin-ham-charge}
	\bar{H} &= \frac{U}{2}\sum_{0 \leq j, k < N_s}  c_{jk} \bar{L}^z_{j+\frac{1}{2}} \bar{L}^z_{k+\frac{1}{2}} + \frac{Y}{2} \sum_{i=0}^{N_s} (\bar{L}_{i+\frac{1}{2}}^z)^2 \nonumber \\
    &- \frac{X}{2} \sum_{i = 1}^{N_s} (\bar{U}_{i-\frac{1}{2}}^+ \bar{U}_{i+\frac{1}{2}}^- + \bar{U}_{i-\frac{1}{2}}^- \bar{U}_{i+\frac{1}{2}}^+) \ ,
\end{align}
where $c_{jk} = N_s - \max\{j, k\}$. Note that Hamiltonian \eqref{eq:any-spin-ham} has $N_s$ sites, while there are $N_s +1$ sites in Hamiltonian \eqref{eq:any-spin-ham-charge}. The former has total charge zero, while the latter has charge conservation symmetry and can represent all charge sectors. Although Hamiltonian \eqref{eq:any-spin-ham-charge} has long range interactions in the first term, a similar form has been used to study the real-time dynamics of the Schwinger model with qubits by coarse graining in time \cite{martinez2016real}. In our case, we focus on Hamiltonian \eqref{eq:any-spin-ham} as it has only nearest-neighbor interactions which allows easy implementations onto optical lattices.

In the following, we used the density matrix renormalization group (DMRG) \cite{PhysRevLett.69.2863,schollwock2011density} for our studies in the continuous-time limit.  The finite DMRG algorithm with matrix product state (MPS) \cite{PhysRevLett.75.3537} optimization was performed using the ITensor C++ library \footnote{Version 2.1.1, http://itensor.org/}. Note that the quantum entanglement comes from the nearest-neighbor interaction in the field representation, while it comes from the nearest-neighbor hopping in the charge representation. In the gapless phase where $X$ is relatively larger than $Y$, the bipartite entanglement entropy for the field representation is much smaller than that of the charge representation, so much smaller bond dimension is needed for the former case in MPS.

\subsection{$H$ with the Polyakov loop}
\label{subsec:imph}

Using the reformulation from Sec. \ref{subsec:ploop}, we can follow the same prescription for taking the continuous-time limit from Sec. \ref{subsec:hamil} and apply it to the Polaykov loop.  This implies taking the same limit for the $P$ operator.  We find,
\begin{equation}
	P \rightarrow 1 + \frac{1}{2(2\kappa_{\tau})} (2(m-m')-1) + \mathcal{O}((2\kappa_{\tau})^{-2})
\end{equation}
This corresponds to an additional term in the quantum Hamiltonian which is located at a single specific site,
and allows us to write,
\begin{equation}
\label{eq:htilde}
	\tilde{H} = H - \frac{Y}{2} (2(L_{i^{*}+1}^{z} - L_{i^{*}}^{z}) - 1)
\end{equation}
where $\tilde{H}$ is the quantum Hamiltonian corresponding to the addition of a static charge at a specific site, $i^{*}$, and $H$ the original quantum Hamiltonian of the Abelian Higgs model.

After taking this limit, it's clear that the energy gap, $\Delta E$, between the ground states of $\tilde{H}$ and $H$ is due to the additional term and the inserted static charge.  This energy gap, when calculated in the low-temperature limit, is the same quantity that is calculated in the isotropic coupling case at large $N_{\tau}$ from Sec. \ref{subsec:deltae}.

In numerical calculations, and to avoid the effects of boundary conditions as much as possible,
if the Polyakov loop is put on the middle link of the square lattice (in the isotropic coupling picture), or equivalently if we remove the ($N_{s}/2$)\textsuperscript{th} term from the $Y$ sum in Hamiltonian \eqref{eq:any-spin-ham} and shift it by one, the Hamiltonian with the Polyakov loop included reads,
\begin{align}
\label{eq:plooph}
	\tilde{H} &= \frac{U}{2} \sum_{i = 1}^{N_s} (L_i^z)^2 + \frac{Y}{2} {\sum_{i \neq \frac{N_s}{2}}} ' (L_{i+1}^z - L_{i}^z)^2 \nonumber \\
    &+ \frac{Y}{2} (L_{\frac{N_s}{2}+1}^z - L_{\frac{N_s}{2}}^z - 1)^2 -X \sum_{i = 1}^{N_s} U_i^x.
\end{align}
By a simple rearranging of terms this can be cast into the form from Eq. \eqref{eq:htilde}.  So adding a Polyakov loop creates a single charge in the neutral system (\emph{i.e.} Eq. \eqref{eq:any-spin-ham}).

 \begin{figure}
 \centering
   \includegraphics[width=0.49\textwidth]{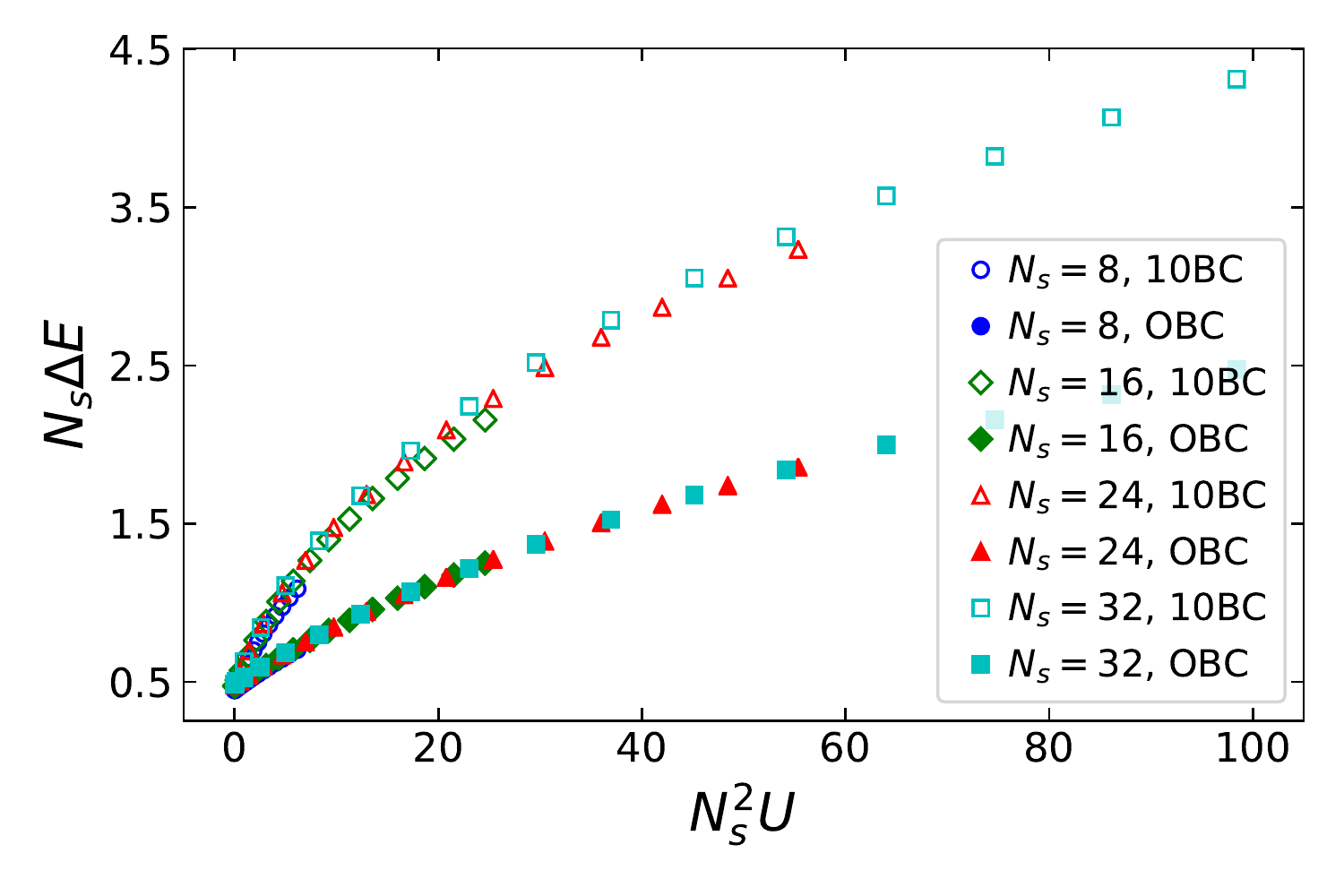}
    \caption{\label{fig:ploop-01bccollapseg2} Data collapse of $N_s \Delta E$ as a function of $N_s^2 U$ with $X = 2$.  The upper curve (open markers) is the data collapse in the case of an external electric field with no Polyakov loop (see Sec. \ref{subsec:contimelim}), and the lower curve (solid markers) is the case of OBC with a Polaykov loop included.  These calculations were done using the DMRG with MPS.}
  \end{figure} 
Just as in the isotropic-coupling picture, the continuous-time limit preserves the collapse of $\Delta E$ across a range of couplings and spatial sizes \cite{Zhang18}.  A glance at Eq. \eqref{eq:ploop-zratio} shows that in the continuous-time limit $\Delta E$ is the difference between the ground state energy eigenvalues of a system with the Polyakov loop included, and one without it.  This can be readily calculated from the DMRG, and the collapse can be seen in Fig. \ref{fig:ploop-01bccollapseg2}.  Even more remarkable is that in an appropriate regime, up to a rescaling by $\kappa$, the $\Delta E$ calculated in the continuous-time limit is equal to the $\Delta E$ calculated in the isotropic-coupling discrete case.  Further details about this can be found in Ref. \cite{Zhang18} where this relationship is proposed as an observable seen in quantum simulations of this model.

\subsection{The $O(2)$ limit}

In our previous work, we discussed the $O(2)$ limit where $U \rightarrow 0$ ($g^2 \rightarrow 0$), and the $\kappa_{\tau} \gg \beta_{pl} \gg \kappa_s$ limit separately, where we used the charge representation and field representation with a three state truncation respectively. Because the $O(2)$ model has a gapless phase where high quantum number states are easily excited, a three-state truncation is far from enough for Hamiltonian \eqref{eq:any-spin-ham}. Here we want to connect these two limits continuously by keeping more states in the truncation. 

\begin{figure}[t]
 \centering
   \includegraphics[width=0.49\textwidth]{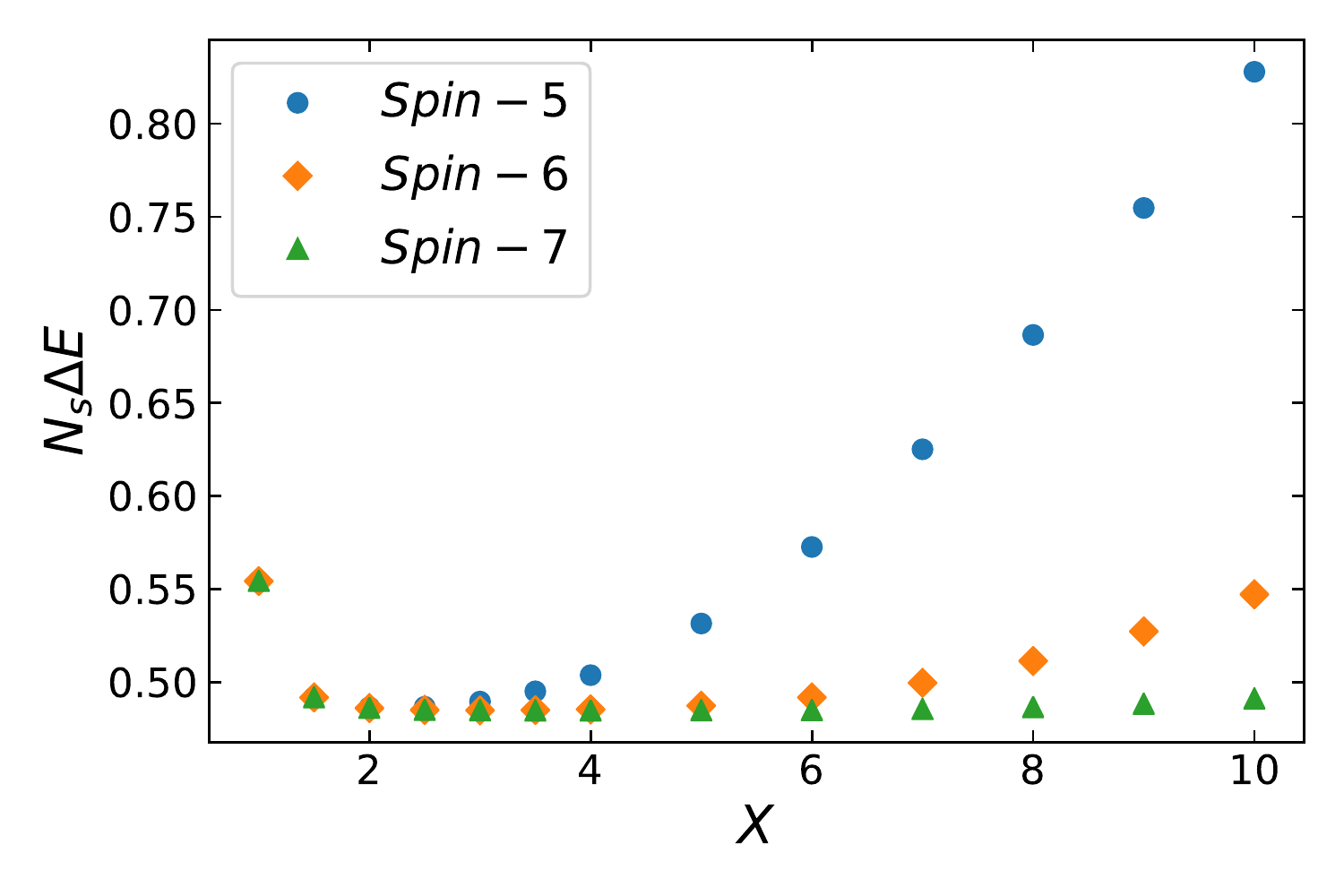}
    \caption{\label{fig:o2fieldX} The enegy gap in the O(2) limit with $N_s = 32$ in field representation as a function of $X$.  These calculations were done using the DMRG with MPS.}
\end{figure}

To test how many states we should keep to simulate the $O(2)$ model using Eq. \eqref{eq:any-spin-ham} with $U = 0$, we gradually increase the number of states and calculate the energy gap between the ground states of Hamiltonians \eqref{eq:any-spin-ham} and \eqref{eq:plooph}.  In these calculations we used units of $Y = 1$ in the Hamiltonian. Studying the data collapse  requires precise values of $N_s \Delta E$ in the $O(2)$ limit. The energy gap of the $O(2)$ model in the KT region scales like $1/N_s$ in a polynomial form, with the coefficient of $1/N_s$ constant, which is verified in Fig. \ref{fig:o2fieldX} where $N_s \Delta E$ is almost constant for a large range of $X \in [2, 10]$ with $N_s = 32$.  We also see that a spin-5 truncation is good enough for $X \approx 2$, and a spin-6 truncation is essentially perfect for $X \in [2, 4]$.
\begin{figure}[t]
\centering
\includegraphics[width=0.49\textwidth]{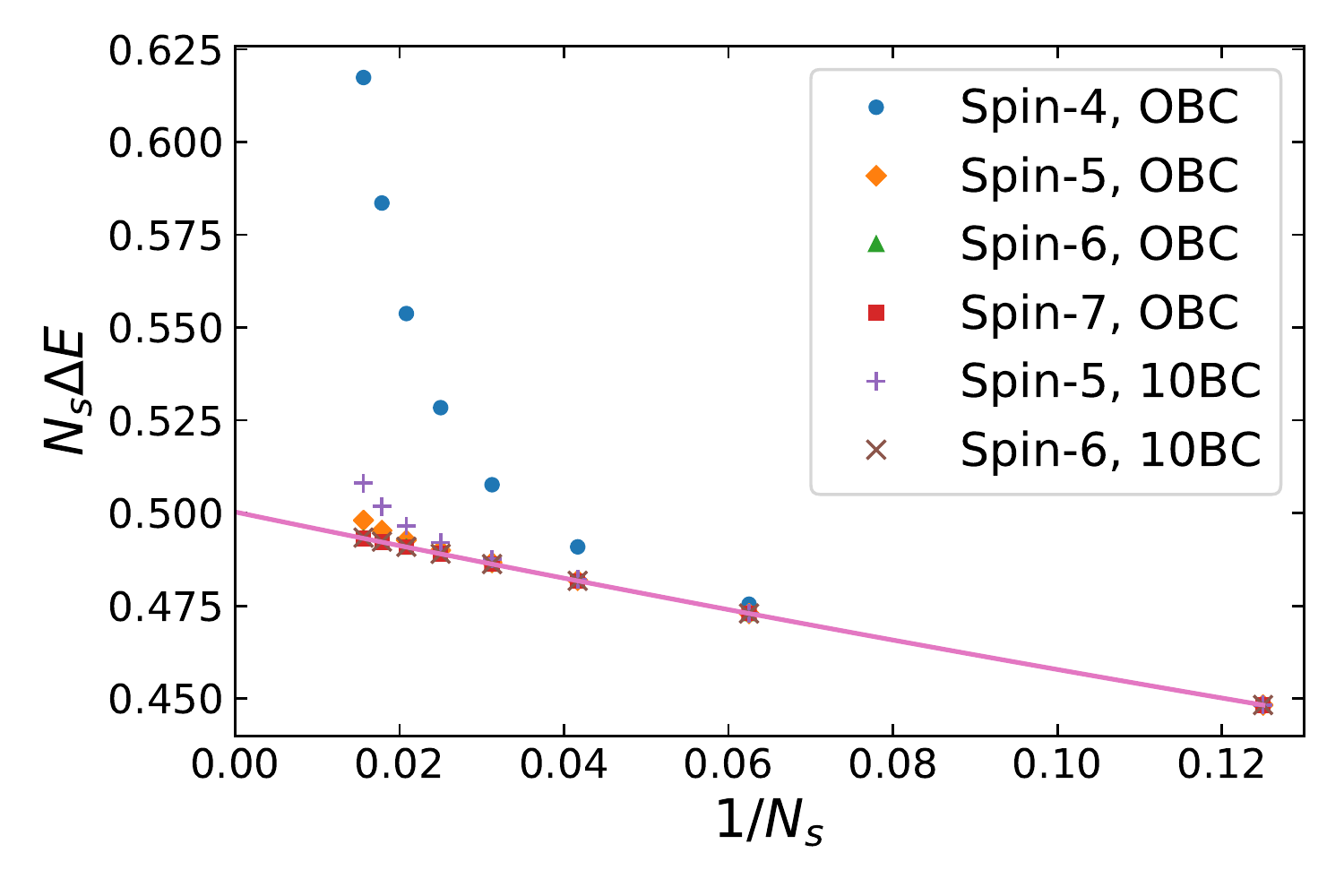}
\caption{\label{fig:o2fieldX21/Ns} The enegy gap in the O(2) limit with $X = 2$ in plaquette quantum number representation as a function of $\frac{1}{N_s}$ for different spin-truncations.  These calculations were done using the DMRG with MPS.}
\end{figure}
\begin{figure}[t]
 \centering
   \includegraphics[width=0.49\textwidth]{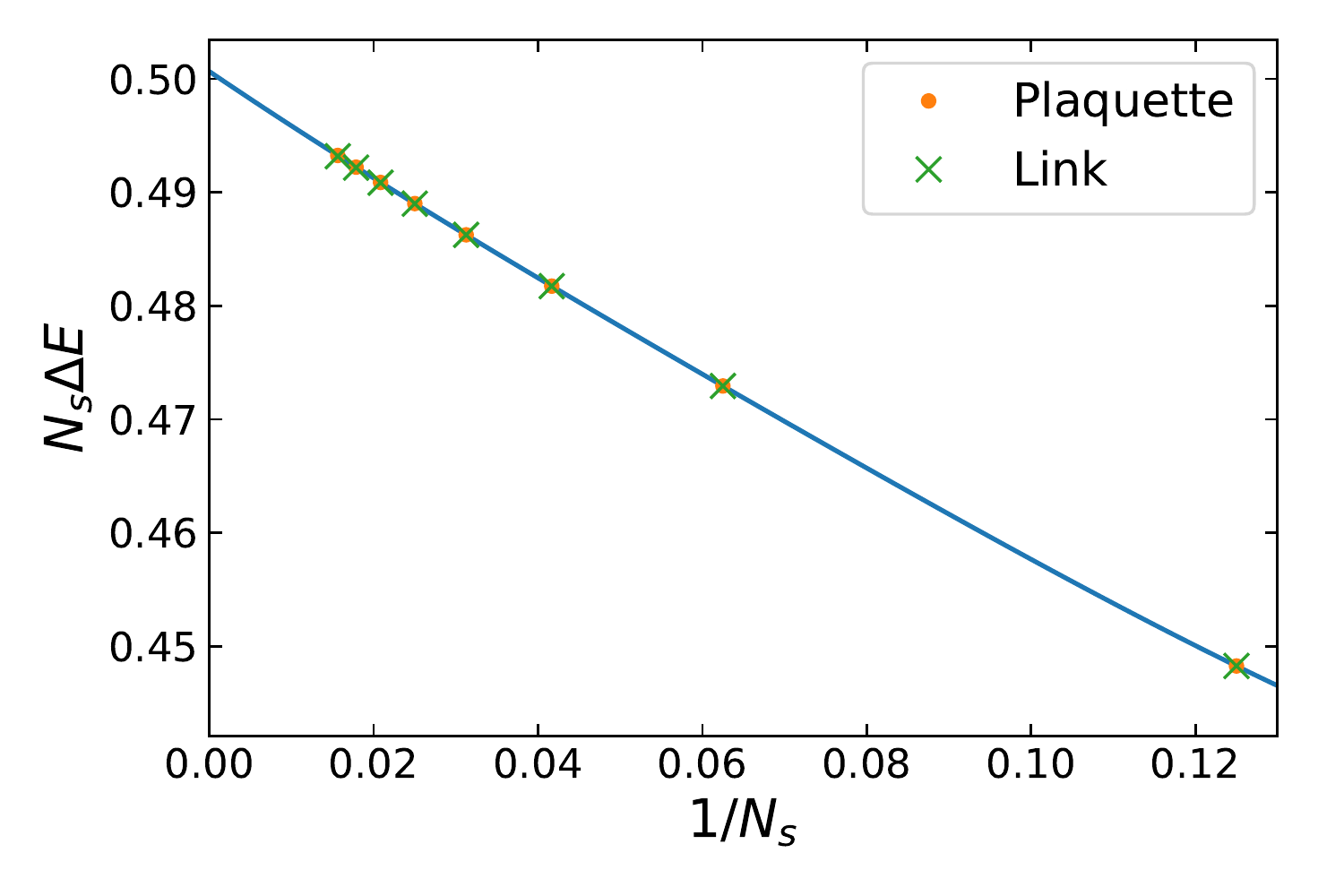}
    \caption{\label{fig:o2chargefield} The enegy gap in the O(2) limit with a spin-6 truncation in the plaquette quantum number representation and link quantum number representation as a function of $1/N_s$. Here $N_s = 8, 16, ..., 64$ with $X = 2$ fixed. These calculations were done with the DMRG using MPS.}
\end{figure}

Finite-size effects also play a role. As shown in Fig. \ref{fig:o2fieldX21/Ns}, at $X = 2$, a spin-5 truncation breaks down after $N_s = 40$, but a spin-6 truncation works perfectly well for $N_s \leq 64$. The extrapolation of $N_s \Delta E$ to the thermodynamic limit gives $N_s \Delta E = 0.50029(8)$, which is consistent with
$Y / 2 = 0.5$ in units of $Y$ in the time continuum limit. By calculating the energy gap between charge-0 and charge-1 sectors with Hamiltonian \eqref{eq:any-spin-ham-charge}, we show that the energy gaps for the two representations of $O(2)$ converge to the same values with a spin-6 truncation.  This can be seen in Fig. \ref{fig:o2chargefield}.\\


\section{Replacing the Polyakov loop by special boundary conditions}
\label{sec:01bc}

\subsection{Isotropic coupling}

By inserting the Polyakov loop, one probes the response of the system to the addition of a single static charge.
For a total charge, $Q$, the $Q \neq 0$ sectors of the theory can be probed by changing the boundary conditions of the system, this is like subjecting the system to an external electric field.

One can consider the pure Abelian Higgs model with a boundary of zeros on one side, and a boundary of ones on the other (10BC) in the field-quantum-number representation.  With the conventions in this text, to put the system in the $Q=-1$ sector it amounts to setting a boundary of ones on the left side of the system, and a boundary of zeros on the right side.  With these boundary conditions (and the now absent Polyakov loop) Gauss' law tells us that there is a total charge of $-1$ across a time slice.  In the tensor language the boundary tensors, in this case the $B$ tensors, are assigned the state $m = 1$ on one side, and $m = 0$ on the other.

The relationship between the situation with the Polyakov loop inserted, and the system with special boundary conditions can be made more clear by a simple example. It consists of sliding the Polyakov loop all the way to the boundary of a system with OBC. We will relate the $Q = -1$ sector (which is identical to the $Q = 1$ sector) to the Polyakov loop.  Take a system where, along a time slice, charge is defined at a link as the difference of the electric field as $n = m_{\text{right}} - m_{\text{left}}$, with $Q = \sum n$.  Then at the link with the Polyakov loop one has $n = m_{\text{right}} - m_{\text{left}} - 1$.  If one slides this all the way to the left-most boundary link, it becomes $n = m_{\text{right}} - 0 - 1 = m_{\text{right}} - 1$.  This is precisely the charge one would find by setting the left boundary quantum numbers to one and removing the Polaykov loop entirely.  By Gauss' law the charge of this system would be $-1$.  Thus, a system with skewed boundary conditions of zeros on one side and ones on the other is equal to a system with OBC with a Polaykov loop pushed to the boundary link.
\begin{figure}[t]
	\includegraphics[width=0.49\textwidth]{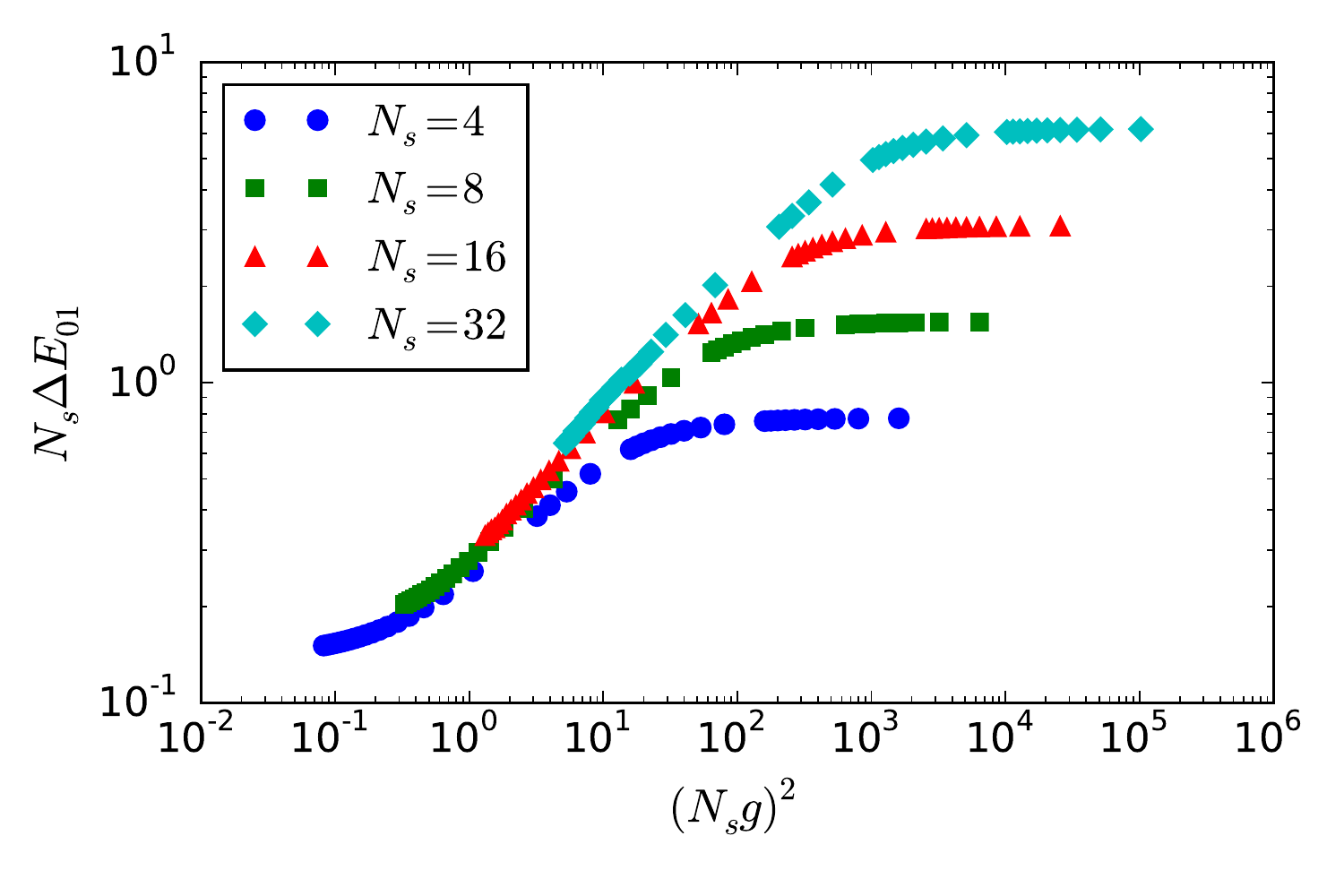}
    \caption{The energy gap between the 01-boundary condition partition function and the OBC partition function in the case of isotropic coupling.  This is for $\kappa = 1.6$ and $D_{\text{bond}} = 41$ for the HOTRG truncation.  Similar to the Polyakov loop gap, for sufficiently small $g$ we see data collapse, and for $g$ large enough we see the collapse breakdown.}
    \label{fig:01bccollapse}
\end{figure}

Since the Polyakov loop is related to a special circumstance of boundary conditions,
we would expect the energy gap from the inclusion of the Polyakov loop to be qualitatively similar to the energy gap between the $Q=1$ and $Q=0$ sectors.
Indeed, a collapse can be found for the energy gap between the 10BC system, and that with OBC.  This is shown in Fig. \ref{fig:01bccollapse}.  Again, there is collapse for sufficiently small $g$, and collapse breaking at large enough $g$.  In addition, $\Delta E_{01} \rightarrow -\ln(t_{1}(2\kappa))$ for large $g$ regardless of $N_{s}$, which is in agreement with the Polyakov loop result.

\subsection{Continuous-time limit}
\label{subsec:contimelim}

Similar to the isotropic coupling case, in the continuous-time limit special boundary conditions can be imposed to probe the $Q \neq 0$ sectors of the theory.  Whereas before, in Sec. \ref{subsec:gen_ham} \& \ref{subsec:imph}, OBC were used (zeros on both spatial ends) and the Polyakov loop was inserted in the center, one can consider the pure model in the absence of the Polyakov loop, and change the left spatial-boundary end-point to one.  By Gauss' law, one can see that this leaves the system with total charge $Q = -1$.  

To implement these boundary conditions in DMRG, one has to imagine there are two additional sites on the chain, one to the left, and one to the right.  The right-side site, say, has quantum number zero (the same as in the typical OBC case), however on the left-side site we assign value one.  To include the effect of the boundary conditions we re-write the Hamiltonian accordingly:
\begin{align}
\label{eq:plooph01}
	H_{10} &= \frac{U}{2} \sum_{i = 1}^{N_s} (L_i^z)^2 +
    \frac{Y}{2} \sum_{i = 1}^{N_{s} - 1} (L_{i+1}^z - L_{i}^z)^2 \nonumber \\
    &+ \frac{Y}{2} (L_{N_s}^z)^2 + \frac{Y}{2} (L_{1}^{z} - 1)^{2} - X \sum_{i = 1}^{N_s} U_i^x.
\end{align}
This is the form of the Hamiltonian in the $Q = -1$ sector, in contrast to \eqref{eq:any-spin-ham} which is the Hamiltonian with zeros on the boundaries (the $Q=0$ sector).  In Fig. \ref{fig:ploop-01bccollapseg2} one can see the data collapse of the energy gap between systems with 10BC and OBC across a range of spatial sizes and gauge couplings.  An understanding of the relationship between the $Q=-1$ Hamiltonian, and the Polyakov loop Hamiltonian can be understood through a change of variables.  This is worked out in Ref. \cite{Zhang18} where the two systems are related by a linear potential term.

\section{Conclusions}
\label{sec:conclusion}

In this paper we have explored the inclusion of the Polyakov loop into the compact Abelian Higgs model in 1+1 dimensions.  In our study we compared the TRG, MC, and the DMRG methods and found excellent agreement between them.  It was found that the Polaykov loop is related to the energy gap between charge-sectors of the Abelian Higgs model and that this energy gap exhibits universal finite-size scaling behavior.  The scaling of the energy gap was studied in both the fully discrete lattice system and the continuous-time quantum limit of this model, and it was found that the universal behavior of the energy gap survives this limit.  In addition, special boundary conditions were able to reproduce similar features of the data collapse found from the Polyakov loop which provides an alternate method to study this energy gap.

The results in this paper give support to Ref. \cite{Zhang18} where it is proposed to use the 2D Abelian Higgs model as a proof of principle model for the case of quantum simulating using cold atoms in optical lattices.  In that reference the energy gap associated with the Polyakov loop is suggested as an observable for the simulation.

An interesting feature found in our study is that at finite space-time volume and weak gauge coupling,  there is a sudden increase of the Polyakov loop when we increase
$\kappa$ beyond the critical value corresponding to the KT transition.

\begin{acknowledgments}

We thank J. Zeiher for valuable conversations. 
Part of this work was done while Y.M. visited  the 
GGI in Florence during the workshop ``From Static to Dynamical Gauge Fields with Ultracold Atoms".
This work was supported in part by the U.S. Department of Energy (DOE) under Award Number DE-SC0010113 (YM) and DE-SC0009998 (JUY) and by the NSF under Grant No. DMR-1411345 (SWT). 

\end{acknowledgments}


\begin{thebibliography}{24}%
\makeatletter
\providecommand \@ifxundefined [1]{%
 \@ifx{#1\undefined}
}%
\providecommand \@ifnum [1]{%
 \ifnum #1\expandafter \@firstoftwo
 \else \expandafter \@secondoftwo
 \fi
}%
\providecommand \@ifx [1]{%
 \ifx #1\expandafter \@firstoftwo
 \else \expandafter \@secondoftwo
 \fi
}%
\providecommand \natexlab [1]{#1}%
\providecommand \enquote  [1]{``#1''}%
\providecommand \bibnamefont  [1]{#1}%
\providecommand \bibfnamefont [1]{#1}%
\providecommand \citenamefont [1]{#1}%
\providecommand \href@noop [0]{\@secondoftwo}%
\providecommand \href [0]{\begingroup \@sanitize@url \@href}%
\providecommand \@href[1]{\@@startlink{#1}\@@href}%
\providecommand \@@href[1]{\endgroup#1\@@endlink}%
\providecommand \@sanitize@url [0]{\catcode `\\12\catcode `\$12\catcode
  `\&12\catcode `\#12\catcode `\^12\catcode `\_12\catcode `\%12\relax}%
\providecommand \@@startlink[1]{}%
\providecommand \@@endlink[0]{}%
\providecommand \url  [0]{\begingroup\@sanitize@url \@url }%
\providecommand \@url [1]{\endgroup\@href {#1}{\urlprefix }}%
\providecommand \urlprefix  [0]{URL }%
\providecommand \Eprint [0]{\href }%
\providecommand \doibase [0]{http://dx.doi.org/}%
\providecommand \selectlanguage [0]{\@gobble}%
\providecommand \bibinfo  [0]{\@secondoftwo}%
\providecommand \bibfield  [0]{\@secondoftwo}%
\providecommand \translation [1]{[#1]}%
\providecommand \BibitemOpen [0]{}%
\providecommand \bibitemStop [0]{}%
\providecommand \bibitemNoStop [0]{.\EOS\space}%
\providecommand \EOS [0]{\spacefactor3000\relax}%
\providecommand \BibitemShut  [1]{\csname bibitem#1\endcsname}%
\let\auto@bib@innerbib\@empty
\bibitem [{\citenamefont {Coleman}(1985)}]{coleman}%
  \BibitemOpen
  \bibfield  {author} {\bibinfo {author} {\bibfnamefont {S.}~\bibnamefont
  {Coleman}},\ }\href@noop {} {\emph {\bibinfo {title} {Aspects of Symmetry}}}\
  (\bibinfo  {publisher} {Cambridge University Press},\ \bibinfo {address}
  {Cambridge},\ \bibinfo {year} {1985})\BibitemShut {NoStop}%
\bibitem [{\citenamefont {Martinez}\ \emph {et~al.}(2016)\citenamefont
  {Martinez}, \citenamefont {Muschik}, \citenamefont {Schindler}, \citenamefont
  {Nigg}, \citenamefont {Erhard}, \citenamefont {Heyl}, \citenamefont {Hauke},
  \citenamefont {Dalmonte}, \citenamefont {Monz}, \citenamefont {Zoller} \emph
  {et~al.}}]{martinez2016real}%
  \BibitemOpen
  \bibfield  {author} {\bibinfo {author} {\bibfnamefont {E.~A.}\ \bibnamefont
  {Martinez}}, \bibinfo {author} {\bibfnamefont {C.~A.}\ \bibnamefont
  {Muschik}}, \bibinfo {author} {\bibfnamefont {P.}~\bibnamefont {Schindler}},
  \bibinfo {author} {\bibfnamefont {D.}~\bibnamefont {Nigg}}, \bibinfo {author}
  {\bibfnamefont {A.}~\bibnamefont {Erhard}}, \bibinfo {author} {\bibfnamefont
  {M.}~\bibnamefont {Heyl}}, \bibinfo {author} {\bibfnamefont {P.}~\bibnamefont
  {Hauke}}, \bibinfo {author} {\bibfnamefont {M.}~\bibnamefont {Dalmonte}},
  \bibinfo {author} {\bibfnamefont {T.}~\bibnamefont {Monz}}, \bibinfo {author}
  {\bibfnamefont {P.}~\bibnamefont {Zoller}},  \emph {et~al.},\ }\href@noop {}
  {\bibfield  {journal} {\bibinfo  {journal} {Nature}\ }\textbf {\bibinfo
  {volume} {534}},\ \bibinfo {pages} {516} (\bibinfo {year}
  {2016})}\BibitemShut {NoStop}%
\bibitem [{\citenamefont {Banerjee}\ \emph {et~al.}(2012)\citenamefont
  {Banerjee}, \citenamefont {Dalmonte}, \citenamefont {M\"uller}, \citenamefont
  {Rico}, \citenamefont {Stebler}, \citenamefont {Wiese},\ and\ \citenamefont
  {Zoller}}]{PhysRevLett.109.175302}%
  \BibitemOpen
  \bibfield  {author} {\bibinfo {author} {\bibfnamefont {D.}~\bibnamefont
  {Banerjee}}, \bibinfo {author} {\bibfnamefont {M.}~\bibnamefont {Dalmonte}},
  \bibinfo {author} {\bibfnamefont {M.}~\bibnamefont {M\"uller}}, \bibinfo
  {author} {\bibfnamefont {E.}~\bibnamefont {Rico}}, \bibinfo {author}
  {\bibfnamefont {P.}~\bibnamefont {Stebler}}, \bibinfo {author} {\bibfnamefont
  {U.-J.}\ \bibnamefont {Wiese}}, \ and\ \bibinfo {author} {\bibfnamefont
  {P.}~\bibnamefont {Zoller}},\ }\href {\doibase
  10.1103/PhysRevLett.109.175302} {\bibfield  {journal} {\bibinfo  {journal}
  {Phys. Rev. Lett.}\ }\textbf {\bibinfo {volume} {109}},\ \bibinfo {pages}
  {175302} (\bibinfo {year} {2012})}\BibitemShut {NoStop}%
\bibitem [{\citenamefont {Kasper}\ \emph {et~al.}(2017)\citenamefont {Kasper},
  \citenamefont {Hebenstreit}, \citenamefont {Jendrzejewski}, \citenamefont
  {Oberthaler},\ and\ \citenamefont {Berges}}]{Kasper:2017}%
  \BibitemOpen
  \bibfield  {author} {\bibinfo {author} {\bibfnamefont {V.}~\bibnamefont
  {Kasper}}, \bibinfo {author} {\bibfnamefont {F.}~\bibnamefont {Hebenstreit}},
  \bibinfo {author} {\bibfnamefont {F.}~\bibnamefont {Jendrzejewski}}, \bibinfo
  {author} {\bibfnamefont {M.~K.}\ \bibnamefont {Oberthaler}}, \ and\ \bibinfo
  {author} {\bibfnamefont {J.}~\bibnamefont {Berges}},\ }\href
  {http://stacks.iop.org/1367-2630/19/i=2/a=023030} {\bibfield  {journal}
  {\bibinfo  {journal} {New Journal of Physics}\ }\textbf {\bibinfo {volume}
  {19}},\ \bibinfo {pages} {023030} (\bibinfo {year} {2017})}\BibitemShut
  {NoStop}%
\bibitem [{\citenamefont {Klco}\ \emph {et~al.}(2018)\citenamefont {Klco},
  \citenamefont {Dumitrescu}, \citenamefont {McCaskey}, \citenamefont {Morris},
  \citenamefont {Pooser}, \citenamefont {Sanz}, \citenamefont {Solano},
  \citenamefont {Lougovski},\ and\ \citenamefont {Savage}}]{Klco:2018kyo}%
  \BibitemOpen
  \bibfield  {author} {\bibinfo {author} {\bibfnamefont {N.}~\bibnamefont
  {Klco}}, \bibinfo {author} {\bibfnamefont {E.~F.}\ \bibnamefont
  {Dumitrescu}}, \bibinfo {author} {\bibfnamefont {A.~J.}\ \bibnamefont
  {McCaskey}}, \bibinfo {author} {\bibfnamefont {T.~D.}\ \bibnamefont
  {Morris}}, \bibinfo {author} {\bibfnamefont {R.~C.}\ \bibnamefont {Pooser}},
  \bibinfo {author} {\bibfnamefont {M.}~\bibnamefont {Sanz}}, \bibinfo {author}
  {\bibfnamefont {E.}~\bibnamefont {Solano}}, \bibinfo {author} {\bibfnamefont
  {P.}~\bibnamefont {Lougovski}}, \ and\ \bibinfo {author} {\bibfnamefont
  {M.~J.}\ \bibnamefont {Savage}},\ }\href@noop {} {\enquote {\bibinfo {title}
  {Quantum-classical computations of schwinger model dynamics using quantum
  computers},}\ } (\bibinfo {year} {2018}),\ \bibinfo {note}
  {{a}rxiv:1803.03326}\BibitemShut {NoStop}%
\bibitem [{\citenamefont {Kuno}\ \emph {et~al.}(2015)\citenamefont {Kuno},
  \citenamefont {Kasamatsu}, \citenamefont {Takahashi}, \citenamefont
  {Ichinose},\ and\ \citenamefont {Matsui}}]{kuno2015real}%
  \BibitemOpen
  \bibfield  {author} {\bibinfo {author} {\bibfnamefont {Y.}~\bibnamefont
  {Kuno}}, \bibinfo {author} {\bibfnamefont {K.}~\bibnamefont {Kasamatsu}},
  \bibinfo {author} {\bibfnamefont {Y.}~\bibnamefont {Takahashi}}, \bibinfo
  {author} {\bibfnamefont {I.}~\bibnamefont {Ichinose}}, \ and\ \bibinfo
  {author} {\bibfnamefont {T.}~\bibnamefont {Matsui}},\ }\href@noop {}
  {\bibfield  {journal} {\bibinfo  {journal} {New J. Phys.}\ }\textbf {\bibinfo
  {volume} {17}},\ \bibinfo {pages} {063005} (\bibinfo {year}
  {2015})}\BibitemShut {NoStop}%
\bibitem [{\citenamefont {Kuno}\ \emph {et~al.}(2017)\citenamefont {Kuno},
  \citenamefont {Sakane}, \citenamefont {Kasamatsu}, \citenamefont {Ichinose},\
  and\ \citenamefont {Matsui}}]{kuno2017quantum}%
  \BibitemOpen
  \bibfield  {author} {\bibinfo {author} {\bibfnamefont {Y.}~\bibnamefont
  {Kuno}}, \bibinfo {author} {\bibfnamefont {S.}~\bibnamefont {Sakane}},
  \bibinfo {author} {\bibfnamefont {K.}~\bibnamefont {Kasamatsu}}, \bibinfo
  {author} {\bibfnamefont {I.}~\bibnamefont {Ichinose}}, \ and\ \bibinfo
  {author} {\bibfnamefont {T.}~\bibnamefont {Matsui}},\ }\href@noop {}
  {\bibfield  {journal} {\bibinfo  {journal} {Phys. Rev. D}\ }\textbf {\bibinfo
  {volume} {95}},\ \bibinfo {pages} {094507} (\bibinfo {year}
  {2017})}\BibitemShut {NoStop}%
\bibitem [{\citenamefont {Gonzalez-Cuadra}\ \emph {et~al.}(2017)\citenamefont
  {Gonzalez-Cuadra}, \citenamefont {Zohar},\ and\ \citenamefont
  {Cirac}}]{Cuadra2017}%
  \BibitemOpen
  \bibfield  {author} {\bibinfo {author} {\bibfnamefont {D.}~\bibnamefont
  {Gonzalez-Cuadra}}, \bibinfo {author} {\bibfnamefont {E.}~\bibnamefont
  {Zohar}}, \ and\ \bibinfo {author} {\bibfnamefont {J.~I.}\ \bibnamefont
  {Cirac}},\ }\href@noop {} {\bibfield  {journal} {\bibinfo  {journal} {New J.
  Phys.}\ }\textbf {\bibinfo {volume} {19}},\ \bibinfo {pages} {063038}
  (\bibinfo {year} {2017})}\BibitemShut {NoStop}%
\bibitem [{\citenamefont {Xie}\ \emph {et~al.}(2012)\citenamefont {Xie},
  \citenamefont {Chen}, \citenamefont {Qin}, \citenamefont {Zhu}, \citenamefont
  {Yang},\ and\ \citenamefont {Xiang}}]{PhysRevB.86.045139}%
  \BibitemOpen
  \bibfield  {author} {\bibinfo {author} {\bibfnamefont {Z.~Y.}\ \bibnamefont
  {Xie}}, \bibinfo {author} {\bibfnamefont {J.}~\bibnamefont {Chen}}, \bibinfo
  {author} {\bibfnamefont {M.~P.}\ \bibnamefont {Qin}}, \bibinfo {author}
  {\bibfnamefont {J.~W.}\ \bibnamefont {Zhu}}, \bibinfo {author} {\bibfnamefont
  {L.~P.}\ \bibnamefont {Yang}}, \ and\ \bibinfo {author} {\bibfnamefont
  {T.}~\bibnamefont {Xiang}},\ }\href {\doibase 10.1103/PhysRevB.86.045139}
  {\bibfield  {journal} {\bibinfo  {journal} {Phys. Rev. B}\ }\textbf {\bibinfo
  {volume} {86}},\ \bibinfo {pages} {045139} (\bibinfo {year}
  {2012})}\BibitemShut {NoStop}%
\bibitem [{\citenamefont {Meurice}(2013)}]{prb87}%
  \BibitemOpen
  \bibfield  {author} {\bibinfo {author} {\bibfnamefont {Y.}~\bibnamefont
  {Meurice}},\ }\href {\doibase 10.1103/PhysRevB.87.064422} {\bibfield
  {journal} {\bibinfo  {journal} {Phys. Rev.}\ }\textbf {\bibinfo {volume}
  {B87}},\ \bibinfo {pages} {064422} (\bibinfo {year} {2013})},\ \Eprint
  {http://arxiv.org/abs/1211.3675} {arXiv:1211.3675 [hep-lat]} \BibitemShut
  {NoStop}%
\bibitem [{\citenamefont {Liu}\ \emph {et~al.}(2013)\citenamefont {Liu},
  \citenamefont {Meurice}, \citenamefont {Qin}, \citenamefont {Unmuth-Yockey},
  \citenamefont {Xiang}, \citenamefont {Xie}, \citenamefont {Yu},\ and\
  \citenamefont {Zou}}]{prd88}%
  \BibitemOpen
  \bibfield  {author} {\bibinfo {author} {\bibfnamefont {Y.}~\bibnamefont
  {Liu}}, \bibinfo {author} {\bibfnamefont {Y.}~\bibnamefont {Meurice}},
  \bibinfo {author} {\bibfnamefont {M.~P.}\ \bibnamefont {Qin}}, \bibinfo
  {author} {\bibfnamefont {J.}~\bibnamefont {Unmuth-Yockey}}, \bibinfo {author}
  {\bibfnamefont {T.}~\bibnamefont {Xiang}}, \bibinfo {author} {\bibfnamefont
  {Z.~Y.}\ \bibnamefont {Xie}}, \bibinfo {author} {\bibfnamefont {J.~F.}\
  \bibnamefont {Yu}}, \ and\ \bibinfo {author} {\bibfnamefont {H.}~\bibnamefont
  {Zou}},\ }\href {\doibase 10.1103/PhysRevD.88.056005} {\bibfield  {journal}
  {\bibinfo  {journal} {Phys. Rev.}\ }\textbf {\bibinfo {volume} {D88}},\
  \bibinfo {pages} {056005} (\bibinfo {year} {2013})}\BibitemShut {NoStop}%
\bibitem [{\citenamefont {Denbleyker}\ \emph {et~al.}(2014)\citenamefont
  {Denbleyker}, \citenamefont {Liu}, \citenamefont {Meurice}, \citenamefont
  {Qin}, \citenamefont {Xiang}, \citenamefont {Xie}, \citenamefont {Yu},\ and\
  \citenamefont {Zou}}]{prd89}%
  \BibitemOpen
  \bibfield  {author} {\bibinfo {author} {\bibfnamefont {A.}~\bibnamefont
  {Denbleyker}}, \bibinfo {author} {\bibfnamefont {Y.}~\bibnamefont {Liu}},
  \bibinfo {author} {\bibfnamefont {Y.}~\bibnamefont {Meurice}}, \bibinfo
  {author} {\bibfnamefont {M.~P.}\ \bibnamefont {Qin}}, \bibinfo {author}
  {\bibfnamefont {T.}~\bibnamefont {Xiang}}, \bibinfo {author} {\bibfnamefont
  {Z.~Y.}\ \bibnamefont {Xie}}, \bibinfo {author} {\bibfnamefont {J.~F.}\
  \bibnamefont {Yu}}, \ and\ \bibinfo {author} {\bibfnamefont {H.}~\bibnamefont
  {Zou}},\ }\href {\doibase 10.1103/PhysRevD.89.016008} {\bibfield  {journal}
  {\bibinfo  {journal} {Phys. Rev.}\ }\textbf {\bibinfo {volume} {D89}},\
  \bibinfo {pages} {016008} (\bibinfo {year} {2014})},\ \Eprint
  {http://arxiv.org/abs/1309.6623} {arXiv:1309.6623 [hep-lat]} \BibitemShut
  {NoStop}%
\bibitem [{\citenamefont {Yu}\ \emph {et~al.}(2014)\citenamefont {Yu},
  \citenamefont {Xie}, \citenamefont {Meurice}, \citenamefont {Liu},
  \citenamefont {Denbleyker}, \citenamefont {Zou}, \citenamefont {Qin},\ and\
  \citenamefont {Chen}}]{pre89}%
  \BibitemOpen
  \bibfield  {author} {\bibinfo {author} {\bibfnamefont {J.~F.}\ \bibnamefont
  {Yu}}, \bibinfo {author} {\bibfnamefont {Z.~Y.}\ \bibnamefont {Xie}},
  \bibinfo {author} {\bibfnamefont {Y.}~\bibnamefont {Meurice}}, \bibinfo
  {author} {\bibfnamefont {Y.}~\bibnamefont {Liu}}, \bibinfo {author}
  {\bibfnamefont {A.}~\bibnamefont {Denbleyker}}, \bibinfo {author}
  {\bibfnamefont {H.}~\bibnamefont {Zou}}, \bibinfo {author} {\bibfnamefont
  {M.~P.}\ \bibnamefont {Qin}}, \ and\ \bibinfo {author} {\bibfnamefont
  {J.}~\bibnamefont {Chen}},\ }\href {\doibase 10.1103/PhysRevE.89.013308}
  {\bibfield  {journal} {\bibinfo  {journal} {Phys. Rev.}\ }\textbf {\bibinfo
  {volume} {E89}},\ \bibinfo {pages} {013308} (\bibinfo {year} {2014})},\
  \Eprint {http://arxiv.org/abs/1309.4963} {arXiv:1309.4963
  [cond-mat.stat-mech]} \BibitemShut {NoStop}%
\bibitem [{\citenamefont {Gattringer}\ \emph {et~al.}(2015)\citenamefont
  {Gattringer}, \citenamefont {Kloiber},\ and\ \citenamefont
  {M\"uller-Preussker}}]{PhysRevD.92.114508}%
  \BibitemOpen
  \bibfield  {author} {\bibinfo {author} {\bibfnamefont {C.}~\bibnamefont
  {Gattringer}}, \bibinfo {author} {\bibfnamefont {T.}~\bibnamefont {Kloiber}},
  \ and\ \bibinfo {author} {\bibfnamefont {M.}~\bibnamefont
  {M\"uller-Preussker}},\ }\href {\doibase 10.1103/PhysRevD.92.114508}
  {\bibfield  {journal} {\bibinfo  {journal} {Phys. Rev. D}\ }\textbf {\bibinfo
  {volume} {92}},\ \bibinfo {pages} {114508} (\bibinfo {year}
  {2015})}\BibitemShut {NoStop}%
\bibitem [{\citenamefont {Bazavov}\ \emph {et~al.}(2015)\citenamefont
  {Bazavov}, \citenamefont {Meurice}, \citenamefont {Tsai}, \citenamefont
  {Unmuth-Yockey},\ and\ \citenamefont {Zhang}}]{PhysRevD.92.076003}%
  \BibitemOpen
  \bibfield  {author} {\bibinfo {author} {\bibfnamefont {A.}~\bibnamefont
  {Bazavov}}, \bibinfo {author} {\bibfnamefont {Y.}~\bibnamefont {Meurice}},
  \bibinfo {author} {\bibfnamefont {S.-W.}\ \bibnamefont {Tsai}}, \bibinfo
  {author} {\bibfnamefont {J.}~\bibnamefont {Unmuth-Yockey}}, \ and\ \bibinfo
  {author} {\bibfnamefont {J.}~\bibnamefont {Zhang}},\ }\href {\doibase
  10.1103/PhysRevD.92.076003} {\bibfield  {journal} {\bibinfo  {journal} {Phys.
  Rev. D}\ }\textbf {\bibinfo {volume} {92}},\ \bibinfo {pages} {076003}
  (\bibinfo {year} {2015})}\BibitemShut {NoStop}%
\bibitem [{\citenamefont {Zeiher}\ \emph {et~al.}(2016)\citenamefont {Zeiher},
  \citenamefont {van Bijnen}, \citenamefont {Schau{\ss}}, \citenamefont {Hild},
  \citenamefont {Choi}, \citenamefont {Pohl}, \citenamefont {Bloch},\ and\
  \citenamefont {Gross}}]{Zeiher2016}%
  \BibitemOpen
  \bibfield  {author} {\bibinfo {author} {\bibfnamefont {J.}~\bibnamefont
  {Zeiher}}, \bibinfo {author} {\bibfnamefont {R.}~\bibnamefont {van Bijnen}},
  \bibinfo {author} {\bibfnamefont {P.}~\bibnamefont {Schau{\ss}}}, \bibinfo
  {author} {\bibfnamefont {S.}~\bibnamefont {Hild}}, \bibinfo {author}
  {\bibfnamefont {J.-y.}\ \bibnamefont {Choi}}, \bibinfo {author}
  {\bibfnamefont {T.}~\bibnamefont {Pohl}}, \bibinfo {author} {\bibfnamefont
  {I.}~\bibnamefont {Bloch}}, \ and\ \bibinfo {author} {\bibfnamefont
  {C.}~\bibnamefont {Gross}},\ }\href {\doibase 10.1038/nphys3835} {\bibfield
  {journal} {\bibinfo  {journal} {Nat. Phys.}\ }\textbf {\bibinfo {volume}
  {12}},\ \bibinfo {pages} {1095} (\bibinfo {year} {2016})}\BibitemShut
  {NoStop}%
\bibitem [{\citenamefont {Zhang}\ \emph {et~al.}(2018)\citenamefont {Zhang},
  \citenamefont {Unmuth-Yockey}, \citenamefont {Bazavov}, \citenamefont
  {Tsai},\ and\ \citenamefont {Meurice}}]{Zhang18}%
  \BibitemOpen
  \bibfield  {author} {\bibinfo {author} {\bibfnamefont {J.}~\bibnamefont
  {Zhang}}, \bibinfo {author} {\bibfnamefont {J.}~\bibnamefont
  {Unmuth-Yockey}}, \bibinfo {author} {\bibfnamefont {A.}~\bibnamefont
  {Bazavov}}, \bibinfo {author} {\bibfnamefont {S.~W.}\ \bibnamefont {Tsai}}, \
  and\ \bibinfo {author} {\bibfnamefont {Y.}~\bibnamefont {Meurice}},\
  }\href@noop {} {\  (\bibinfo {year} {2018})},\ \Eprint
  {http://arxiv.org/abs/1803.11166} {arXiv:1803.11166 [hep-lat]} \BibitemShut
  {NoStop}%
\bibitem [{\citenamefont {Savit}(1980)}]{RevModPhys.52.453}%
  \BibitemOpen
  \bibfield  {author} {\bibinfo {author} {\bibfnamefont {R.}~\bibnamefont
  {Savit}},\ }\href {\doibase 10.1103/RevModPhys.52.453} {\bibfield  {journal}
  {\bibinfo  {journal} {Rev. Mod. Phys.}\ }\textbf {\bibinfo {volume} {52}},\
  \bibinfo {pages} {453} (\bibinfo {year} {1980})}\BibitemShut {NoStop}%
\bibitem [{\citenamefont {Montvay}\ and\ \citenamefont
  {Munster}(1997)}]{Montvay:1994cy}%
  \BibitemOpen
  \bibfield  {author} {\bibinfo {author} {\bibfnamefont {I.}~\bibnamefont
  {Montvay}}\ and\ \bibinfo {author} {\bibfnamefont {G.}~\bibnamefont
  {Munster}},\ }\href {\doibase 10.1017/CBO9780511470783} {\emph {\bibinfo
  {title} {{Quantum fields on a lattice}}}},\ Cambridge Monographs on
  Mathematical Physics\ (\bibinfo  {publisher} {Cambridge University Press},\
  \bibinfo {year} {1997})\BibitemShut {NoStop}%
\bibitem [{\citenamefont {POPPITZ}\ and\ \citenamefont
  {SHANG}(2008)}]{doi:10.1142/S0217751X08041281}%
  \BibitemOpen
  \bibfield  {author} {\bibinfo {author} {\bibfnamefont {E.}~\bibnamefont
  {POPPITZ}}\ and\ \bibinfo {author} {\bibfnamefont {Y.}~\bibnamefont
  {SHANG}},\ }\href {\doibase 10.1142/S0217751X08041281} {\bibfield  {journal}
  {\bibinfo  {journal} {International Journal of Modern Physics A}\ }\textbf
  {\bibinfo {volume} {23}},\ \bibinfo {pages} {4545} (\bibinfo {year}
  {2008})},\ \Eprint
  {http://arxiv.org/abs/https://doi.org/10.1142/S0217751X08041281}
  {https://doi.org/10.1142/S0217751X08041281} \BibitemShut {NoStop}%
\bibitem [{\citenamefont {White}(1992)}]{PhysRevLett.69.2863}%
  \BibitemOpen
  \bibfield  {author} {\bibinfo {author} {\bibfnamefont {S.~R.}\ \bibnamefont
  {White}},\ }\href {\doibase 10.1103/PhysRevLett.69.2863} {\bibfield
  {journal} {\bibinfo  {journal} {Phys. Rev. Lett.}\ }\textbf {\bibinfo
  {volume} {69}},\ \bibinfo {pages} {2863} (\bibinfo {year}
  {1992})}\BibitemShut {NoStop}%
\bibitem [{\citenamefont {Schollw{\"o}ck}(2011)}]{schollwock2011density}%
  \BibitemOpen
  \bibfield  {author} {\bibinfo {author} {\bibfnamefont {U.}~\bibnamefont
  {Schollw{\"o}ck}},\ }\href@noop {} {\bibfield  {journal} {\bibinfo  {journal}
  {Ann. Phys.}\ }\textbf {\bibinfo {volume} {326}},\ \bibinfo {pages} {96}
  (\bibinfo {year} {2011})}\BibitemShut {NoStop}%
\bibitem [{\citenamefont {\"Ostlund}\ and\ \citenamefont
  {Rommer}(1995)}]{PhysRevLett.75.3537}%
  \BibitemOpen
  \bibfield  {author} {\bibinfo {author} {\bibfnamefont {S.}~\bibnamefont
  {\"Ostlund}}\ and\ \bibinfo {author} {\bibfnamefont {S.}~\bibnamefont
  {Rommer}},\ }\href {\doibase 10.1103/PhysRevLett.75.3537} {\bibfield
  {journal} {\bibinfo  {journal} {Phys. Rev. Lett.}\ }\textbf {\bibinfo
  {volume} {75}},\ \bibinfo {pages} {3537} (\bibinfo {year}
  {1995})}\BibitemShut {NoStop}%
\bibitem [{Note1()}]{Note1}%
  \BibitemOpen
  \bibinfo {note} {Version 2.1.1, http://itensor.org/}\BibitemShut {NoStop}%
\end{thebibliography}
%

\end{document}